\documentclass[a4paper,11pt]{article}
\pdfoutput=1 

\usepackage{jheppub} 

\usepackage[T1]{fontenc} 
\usepackage{mathtools}
\usepackage{url}

\usepackage[usenames,dvipsnames]{xcolor}
 \usepackage{musicography}
\usepackage{empheq}
  \usepackage{mathrsfs}
 
\usepackage{bm}
\usepackage{verbatim}
 
 \usepackage{makecell}
 \usepackage{float}
 \usepackage{hyperref}
\usepackage[normalem]{ulem}

 \usepackage[makeroom]{cancel}
 \usepackage{tikz}
 \usetikzlibrary{shapes.misc}

 \tikzset{cross/.style={cross out, draw=black, minimum size=2*(#1-\pgflinewidth), inner sep=0pt, outer sep=0pt},
cross/.default={2pt}}

 \newcommand{\bv}{ \begin{verbatim}}

\usepackage[pagewise]{lineno}

\newcommand{\be}{\begin{equation}}
\newcommand{\ee}{\end{equation}}
\newcommand{\bpm}{\begin{pmatrix}}
\newcommand{\epm}{\end{pmatrix}}

\newcommand{\beqn}{\begin{eqnarray}}
\newcommand{\eeqn}{\end{eqnarray}}

\usepackage{slashed}

\newcommand{\contract}[1]{\text{\Large$:$}#1\text{\Large $:$ } }

\title{Celestial holography from Chiral strings}

\author{Hare Krishna} \author{and Yu-Ping Wang}
\affiliation{C.N. Yang Institute for Theoretical Physics, Stony Brook University,\\ Stony Brook, NY 11794, USA}

\emailAdd{harekrishna.harekrishna@stonybrook.edu}\emailAdd{yu-ping.wang@stonybrook.edu}

\abstract{ 
    In this paper, we studied the relationship between celestial holography and chiral strings. Chiral strings differ from the usual string theory by a change of boundary conditions on the string propagators. It is shown that chiral strings would reproduce graviton amplitudes and could serve as an alternative description of Einstein's gravity. Celestial holography is a proposed duality between gravity in asymptotically flat space-time and a CFT living on its conformal boundary. Since both are CFT descriptions of gravity, we investigate the potential connection between these two formalisms.

    In this paper, we find that both the energetic as well as conformal soft theorems could be derived from the OPEs of vertex operators of chiral strings. All operators in the CCFT can be described by Mellin transforming the vertex operators in the chiral string theories, and the OPE coefficients of CCFT can also be obtained from the world-sheet description.
}
\begin{document} 
\hfill YITP-SB-2023-25

\maketitle
\flushbottom
\section{Introduction}

An alternative description of gravity amplitudes has been an active research area in the past decade. One approach is to try to find a world-sheet description of such amplitudes. For example, Cachazo, He, and Yuan (CHY) developed a prescription for massless particle amplitudes in arbitrary dimensions, where the amplitude is represented as an integration over moduli space of $n$-punctured sphere   \cite{Cachazo:2013hca, Cachazo:2013gna,  Cachazo:2013iea}.

Later, Mason and Skinner developed a chiral infinite tension analog of string theory which contains only the massless sectors of the string spectrum \cite{Mason:2013sva}.   It is called the ambitwistor strings; It is a world-sheet model with ambitwistor space as its target space.  In particular, the $\mathcal{N} = 2$ version of ambitwistor strings would compute the CHY formula for the NS-NS sector of supergravtity\cite{Berkovits:2013xba}.

In this paper, we shall focus on another description of the particle amplitudes that are related to the two formalism above. It is called the "chiral string" or "left-handed string".
chiral string is proposed by Siegel, where he starts with the usual string theory but changes the boundary condition for its string propagator.

\begin{equation}
G(z, \bar{z}) = \ln z\bar{z} \rightarrow  \ln \frac{z}{\bar{z}},
\end{equation}
and taking the so-called HSZ gauge for the world-sheet coordinate.
\begin{equation}
    \bar{z}\rightarrow \sqrt{1+\beta}\bar{z},\,\,\,\,
 z \rightarrow  \frac{1}{\sqrt{1+\beta}}( z-\beta \bar{z}), \quad \beta \rightarrow \infty.
\end{equation}
The amplitude calculated in this gauge limit will reproduce the delta function in its integrand enforcing the scattering equations just like in the CHY formula and ambitwistor strings. For details, see section \ref{hsz gauge}. 

It was later shown by Huang, Siegel, and Yuan that without using the HSZ gauge, one could reproduce graviton amplitudes directly by relating the chiral string amplitudes with open string amplitudes by a modified KLT relations\cite{Huang:2016bdd}.

\begin{equation}
    M_{\textrm{closed}} = A_{\textrm{open}}^{T} \cdot S  \cdot A_{\textrm{open}} \quad \Rightarrow  \quad M_{\textrm{chiral}} = A_{\textrm{open}}^{T} \cdot S  \cdot \tilde{A}_{\textrm{open}},
\end{equation}
where $  M_{\textrm{closed}}$ and $ M_{\textrm{chiral}}$ are the $n$-point usual closed string amplitude and chiral string amplitude (with the modified propagator) respectively.   $A_{\textrm{open}}$ is a vector of $(n-3)!$ independent open string $n$-point amplitudes, while  $\tilde{A}_{\textrm{open}}$ the same vector but with the sign of string tension $\alpha{}'$  flipped. $S$ is the momentum kernel in KLT relation written as a $(n-3)!\times (n-3)!$ matrix \cite{Bjerrum-Bohr:2010pnr}. In the paper \cite{Huang:2016bdd}, they proved that the modified KLT relation computes the graviton amplitudes since all but a finite number of poles cancel out. We would dive into the details of this formalism in section \ref{intro to chiral string}.\\

There is another approach for the CFT description of graviton scattering in asymptotically flat spacetime. It is usually referred to as celestial holography. This formalism tries to find the analog of $\mathrm{AdS/CFT}$ correspondence but in asymptotically flat spacetime. Celestial holography is a proposed duality between gravity in asymptotically flat space-time and a CFT living on its conformal boundary. Massless particles go out to the null infinity. At each instance of retarded time, there is an asymptotic sphere called the celestial sphere. For a $D$-dimensional asymptotically flat space-time, the celestial sphere is geometrically equivalent to a $D-2$ dimensional sphere. The goal of this program is to find the CFT description (which is proposed to live on the celestial sphere) of scattering in the asymptotically flat spacetime. Celestial holography is a large and rapidly developing area of research, so for a deeper explanation, please see the following surveys and the reference within \cite{Pasterski:2021raf, Raclariu:2021zjz, Strominger:2017zoo,Sen:2017nim,Sen:2017xjn}.\\

Given the two independent descriptions of graviton amplitudes as celestial CFT (CCFT) and chiral string world-sheet theory, we are motivated to study the connections between them (see \cite{Adamo:2021lrv, Adamo:2021zpw, Geyer:2014lca} for world sheet of ambitwistor string and CCFT, and other similar approaches in \cite{Adamo:2014yya, Adamo:2019ipt}).  In this paper, we are mainly focusing on the $D = 4$ gravity tree-level amplitudes,  since in this case the CCFT and chiral strings both lived on the Riemann sphere $S^2$, and the connection between them will become more transparent.

In this paper, we mainly focus on studying the soft theorems of graviton amplitudes. In the celestial CFT formalism, soft theorems could be recast into Ward identities of the asymptotic symmetries, which will become the symmetries of the CCFT.  It is interesting to see how the soft theorems manifest themselves from the chiral string perspective. In fact, it was shown by Higuchi and Kawai that in the usual string theories, one could derive its soft theorems purely from the OPEs of its vertex operators \cite{Higuchi:2018vyu}. In sections \ref{energetic soft theorem} and \ref{conformal soft theorem},  we will reproduce the soft theorems (both energetic and conformal) from chiral strings by studying the OPEs of chiral strings vertex operators. In general, we replaced identities in terms of amplitude with identities in terms of vertex operators. Although some of the identities only make sense when considering them in the correlators with other vertex operators, it makes the structure of these identities and the reason why they hold much clearer.

It is worth to emphasize the big picture of this project. The chiral string is just a new 2-dimensional CFT description of Einstein's gravity; more importantly, the CFT is a free theory made out of free bosons and fermions ( akin to string theory) but with a twist of boundary condition. Therefore, it is natural to ask how could we derive what is known about gravity in terms of this novel description.

We aim to explore how the soft theorem manifests within this novel framework, particularly given its universality at the tree level across various gravity theories, including ambitwistor strings, $N=2$ strings, and standard $N=1$ superstrings.

Here is the outline of our article. 
In section \ref{intro to chiral string}, we review the essential formalism of chiral strings. As an example, we calculated the 4-point chiral string amplitudes and show that it agrees with the graviton amplitude in Einstein's gravity. We have provided details about the spectrum of particles in chiral bosonic strings. We also briefly touch upon chiral strings in the HSZ gauge since we will use them to simplify calculations for the supersymmetric case. In the section \ref{energetic soft theorem}, we found the soft graviton theorem can be obtained using chiral strings. In particular, in section \ref{sec:3-1} we showed that for the bosonic chiral strings, the leading, subleading, and sub-subleading soft theorems arise naturally as the OPEs vertex operators. We also showed the universality of leading and subleading soft theorems in the sense that (1) our derivation works both for usual strings and chiral strings (2) the soft factors from the OPE between graviton vertex operator and an arbitrary vertex operators are the same. In section \ref{soft theorem in superstring hsz}, we calculated the OPEs of super-chiral vertex operators and derived the soft factors up to sub-subleading terms. To simplify calculations, we take the HSZ guage. Although we only calculate the OPEs between two graviton vertex operators, we expect that the universality still holds up to sub-leading terms.  

In section \ref{conformal soft theorem}, we extended our formalism to conformal soft theorems, where we define the ``conformal vertex operators'' by simply Mellin transforming the usual vertex operator. The corresponding conformal soft theorems have been derived. The leading, subleading, and sub-subleading ``soft vertex operators'' are also defined in this section. Their OPEs with other hard vertex operators will give the corresponding soft factors. Then, we have provided a candidate stress tensor for CCFT in section \ref{stress tensor}, and it gives the correct OPE with all other operators in the spectrum. In section \ref{worldsheet and CCFT}, investigate the relationship between the OPEs of chiral strings and the OPEs of CCFT operators. In a quite general setting, we could derive the leading OPEs of CCFT from the OPEs of vertex operators of any world-sheet CFTs. (Which includes the chiral strings, of course.)  We finish by summary and discussions in the section \ref{Discussions}.

The question of unitarity, causality, and analyticity \footnote{The celestial QFT amplitude is $\mathcal{A}(\beta,z)$ is a meromorphic function of $\beta$ having poles only at even integers on the real axis. (For details and the definition of $\beta$ see \cite{Arkani-Hamed:2020gyp}.) } in celestial amplitudes (correlation functions of the CCFT) all can be possibly answered from the world-sheet description of chiral string. By better understanding the connections between CCFT and chiral strings, we could gain more understanding of CCFT itself, and vice versa.\\

\section{Introduction to Chiral string}
\label{intro to chiral string}
Chiral string theory is a world-sheet description of particle theories discovered by Siegel \cite{Siegel:2015axg,  Huang:2016bdd}. In particular, the bosonic chiral string theory will reproduce Einstein's gravity coupled with a massive tensor and a tachyonic tensor, while the super-symmetric chiral string theory will reproduce the spectrum of supergravity. This section will review chiral string theory and how it gives us an alternative way to calculate graviton amplitude. 

The core idea of chiral string is that string propagators are not unique and depend on the boundary condition of the world-sheet. If one alters the boundary condition such that the propagator becomes
\begin{equation}\label{chiralprop}
G(z, \bar{z}) = \ln z\bar{z} \rightarrow  \ln \frac{z}{\bar{z}},
\end{equation}
Amplitudes calculated using this new propagator has only finite resonance states. Thus it can be interpreted as theories with only a finite number of particles. 

In the original chiral string formalism \cite{Siegel:2015axg}, we also take the so-called Hohm-Siegel-Zwiebach (HSZ) gauge. 
\begin{equation}
    \bar{z}\rightarrow \sqrt{1+\beta}\bar{z},\,\,\,\,
 z \rightarrow  \frac{1}{\sqrt{1+\beta}}( z-\beta \bar{z}), \quad \beta \rightarrow \infty.
\end{equation}
If we try to calculate the chiral string amplitude under this limit, we would get delta function term factor enforcing the scattering equation just like the graviton amplitude in CHY formula \cite{Cachazo:2013hca}, or in ambitwistor strings \cite{Mason:2013sva}. It is not explicitly shown that such amplitude will coincide with the chiral string amplitudes without the HSZ gauge, but with the correct prescription of vertex operators, the bosonic 4-point amplitude contains the same finite number of poles as in the ungauged version.  (For details on how to construct the amplitude, see Appendix A of \cite{LipinskiJusinskas:2019cej}. )

Later it was established that we don't need to take this gauge limit if we want to get graviton amplitudes \cite{Huang:2016bdd}.  It was shown that using the modified propagator directly in (\ref{chiralprop}) will reproduce a modified KLT formula, which also gives graviton amplitudes \cite{Kawai:1985xq}. Since using the HSZ gauge simplifies the calculation, we will use it under some circumstances mainly in the discussion of OPE for superstring theory (see section \ref{soft theorem in superstring hsz}). In the next subsection, we are going to demonstrate how chiral string reproduces the graviton four-point amplitudes using the "modified" KLT formula.

\subsection{Amplitudes}
In this section, we will derive the modified KLT formula for chiral strings, and show that it indeed produces the graviton amplitude for the 4-point case.

The calculation for $n$-point chiral string amplitude is the same as for the ordinary strings, it is just that we need to use the modified string propagator. To help us evaluate the integral, we shall do an analytic continuation of the complex world-sheet coordinate $z , \bar{z}$ to left and right moving coordinates $z_{+}, z_{-}$. 

\begin{equation}\label{acont}
\begin{matrix}
    z = \sigma_{1} + i\sigma_{0} \\ 
    \bar{z} = \sigma_{1} - i\sigma_{0} 
\end{matrix}, \quad \Rightarrow \quad
\begin{matrix}
    z  \rightarrow z_{+ }= \sigma_{1} +\sigma_{0} \\ 
    \bar{z} \rightarrow z_{-} = \sigma_{1} - \sigma_{0} 
\end{matrix}
\end{equation}

Note that the propagator (\ref{chiralprop}) has a branch cut, for us to choose the branch consistently, we need to take care of the $i\epsilon$ prescription of the propagator. Such $i\epsilon$ prescription is given by
\begin{equation}\label{acontprop}
    \ln \left(\frac{z}{\bar{z} - i\epsilon/z}\right) \; \rightarrow \;   \ln \left(\frac{z_{+}}{z_{-}} + i\epsilon\right) \;\xRightarrow{\epsilon \rightarrow 0+}\; \ln|z_{+}| - \ln|z_{-}| + i\pi \theta (-z_{+}z_{-}).
\end{equation}
$\theta(x)$ is the step function and we could verify this $i\epsilon$ prescription indeed gives the correct delta function
\begin{equation}
\partial \bar{\partial }G(z, \bar{z}) = \frac{i\epsilon}{(z\bar{z} + i\epsilon)^2} \rightarrow i\pi \delta^{(2)}(z,\bar{z}), \quad \epsilon \rightarrow 0+.
\end{equation}

Recall that the $n$-point tree graviton amplitude is the correlation function between 3 unintegrated vertex operators and $n-3$ integrated vertex operators.
\begin{equation}
    M_{n}(k_{1}, \cdots k_{n}) = \left\langle \; W(k_{1}; 0)W(k_{n-1}; 1)W(k_{n}; \infty) U(k_{n-2})\cdots U(k_{n})\; \right\rangle,
\end{equation}
where,
\begin{equation}
    W(k; z) = c\bar{c}V(k; z), \quad U(k) = \int d^{2}z V(k; z), \quad V(k; z) = e_{\mu\nu}\partial X^{\mu}\bar{\partial}X^{\nu}e^{ik\cdot X}. 
\end{equation}
$e_{\mu\nu}$ is the polarization tensors. In 4-dimensional case, one could decompose it into $\varepsilon_{\mu}\tilde{\varepsilon}_{\nu}$. Note that propagators for the holomorphic and anti-holomorphic ghost are also modified. 
\begin{eqnarray}\label{ghostprop}
        c(z)b(0) \sim \frac{1}{z} , \quad   \tilde{c}(\bar{z})\tilde{b}(0) \sim -\frac{1}{\bar{z}} \nonumber\\
        \gamma(z)\beta(0) \sim \frac{1}{z} , \quad   \tilde{\gamma}(\bar{z})\tilde{\beta}(0) \sim -\frac{1}{\bar{z}}
\end{eqnarray}
In a similar spirit, the OPE of fermions can also be written as
\begin{equation}\label{fermionprop}
      \psi^{\mu}(z)\psi^{\nu}(0) \sim \eta^{\mu\nu}\frac{1}{z} , \quad   \tilde{\psi}^{\mu}(\bar{z})\tilde{\psi}^{\nu}(0) \sim -\eta^{\mu\nu}\frac{1}{\bar{z}} .
\end{equation}
One point to note is that the holomorphic part remains the same but the anti-holomorphic part picks up a negative sign. \\

Calculating the chiral string correlators is the same as usual. We can use the exponentiate trick \cite{Polchinski:1998rq}. where one exponentiated graviton vertex operator.
\begin{equation}
 V(k_{i}; z) \rightarrow e^{ik_{i}\cdot X + \varepsilon_{i} \cdot \partial X + \tilde{\varepsilon}_{i} \cdot \bar{\partial} X.}
\end{equation} 

Using the exponentiated version of vertex operators, one just needs to calculate the contractions between any two exponents, which is much easier. After calculating the correlators of exponentiated vertex operators,  we can extract the original correlators by finding their multi-linear expansions in $\varepsilon_{i}$ and $\tilde{\varepsilon}_{i}$. 

After the analytically continuing (as done in \ref{acont}) we can see that the amplitude factors into integration over left-moving coordinates and the integration over right-moving coordinates, intertwined with some phase factor. We have set $\alpha{}' = 1$ in places that wouldn't cause confusion.
\begin{align}
    M_{n} & \sim \int_{0}^{\infty} \prod_{i =2}^{n -2}dz_{i+} |z_{i+}|^{\frac{1}{2}k_{1}\cdot k_{i}}|1-z_{i+}|^{\frac{1}{2}k_{n-1}\cdot k_{i}}\prod_{i < j < n-1}|z_{ij+}|^{\frac{1}{2}k_{i}\cdot k_{j}}\;C(z_{i+}, k_{i}, \varepsilon_{i}; \alpha{}')\nonumber\\
    \cdot & \int_{0}^{\infty} \prod_{i =2}^{n -2}dz_{i-} |z_{i-}|^{-\frac{1}{2}k_{1}\cdot k_{i}}|1-z_{i-}|^{-\frac{1}{2}k_{n-1}\cdot k_{i}}\prod_{i < j < n-1}|z_{ij-}|^{-\frac{1}{2}k_{i}\cdot k_{j}}\;\overline{C}(z_{i-}, k_{i}, \tilde{\varepsilon}_{i}; -\alpha{}') \nonumber\\
    \cdot  & \;\; \delta\left(\sum^{n}_{i}k_{i}\right) e^{i\pi \Phi(z_{1+}, z_{1-} \cdots z_{n-},  z_{n+})}.
\end{align}
Where $C(z_{i+}, k_{i}, \varepsilon_{i}; \alpha{}')$ and $\overline{C}(z_{i-}, k_{i}, \tilde{\varepsilon}_{i}; -\alpha{}')$  are polynomials in $z_{i\, \pm}$, momentum $k_{i}$ and polarization vectors  $\varepsilon_{i}$, $\tilde{\varepsilon}_{i}$. They are determined by calculating the  contractions between $\varepsilon_{i}\cdot \partial X$ and $e^{ik_{j}\cdot X}$ or contractions between $\varepsilon_{i}\cdot \partial X$ and $\varepsilon_{j}\cdot \partial X$. While the Koba-Nielsen factors come from the contractions between $e^{ik_{i}\cdot X}$ and $e^{ik_{j}\cdot X}$.  Compare this result to the ordinary string theory, we see that $C$ should be the same regardless of usual string theory or chiral string theory, while $\overline{C}$ would change due to the difference in $XX$ OPE, but the difference could be completely compensated by flipping the sign of string tension $\alpha{}'$. That's why we explicitly write down the  $\alpha{}'$  dependence in $C$ and $\overline{C}$. 

The phase factor $\Phi(z_{1}, \cdots z_{n})$, which comes from the propagator (\ref{acontprop}), is 
\begin{equation}
    \Phi(z_{1+}, z_{1-} \cdots z_{n-},  z_{n+}) = \sum_{i < j}^{n}k_{i}\cdot k_{j} \, \theta (-z_{ij+}z_{ij-}).
\end{equation}

Therefore, we can express the chiral string amplitude in terms of (slightly modified) open string amplitude (The modified KLT relation)
\begin{equation}\label{chiralanpfull}
    M_{n}(1, \cdots n) = \sum_{\sigma, \tau}\exp\left[i\pi  \sum_{i, j}^{n}k_{\sigma(i)}\cdot  k_{\tau(j)}\theta_{ij}(\sigma , \tau) \right]A_{n}(\sigma(1, \cdots n ); \alpha{}')A_{n}(\tau(1, \cdots n); -\alpha{}').
\end{equation}
 The summation over $\sigma, \tau$ are all permutations of $\{1, \cdots, n\}$ where the order of $1, n-1, n$ are fixed. The phase factor $\theta_{ij}(\sigma, \tau)$ is defined as
\begin{equation}
    \theta_{ij}(\sigma , \tau) =    
    \begin{cases}
        1 & \text{If the order of $\sigma(i), \sigma(j)$ and $\tau(i), \tau(j)$ are opposite.} \\
        0 & \text{If the order of $\sigma(i), \sigma(j)$ and $\tau(i), \tau(j)$ are same.} 
    \end{cases}
\end{equation}
Note that, unlike the usual KLT relations, one of the open string amplitudes has the sign of the tension $\alpha{}'$ flipped, which comes from the modified propagator (\ref{acontprop}). In the next subsection, we will give an explicit example which will make it more explicit.

\subsubsection{Four-point amplitude}
The modified KLT relation will reproduce the graviton amplitude. This is proven in \cite{Huang:2016bdd} order by order in $\alpha{}'$. Although it requires some conjecture about the structures of the open string amplitude (See \cite{Huang:2016bdd, Schlotterer:2012ny}).

For the four-point chiral amplitude, this can be verified directly. Using (\ref{chiralanpfull}), we have
\begin{align}\label{4ptraw}
    &M(1, 2, 3, 4) = \nonumber \\
    & A_{R}(2, 1, 3, 4)\left[ A_{L}(2, 1, 3, 4) +  e^{i\pi k_{1}\cdot k_{2}}A_{L}( 1, 2, 3, 4)   +   e^{i\pi  k_{2}\cdot( k_{1} + k_{3})}A_{L}(1, 3, 2,  4) \right]\nonumber\\
    &A_{R}(1, 2,  3, 4)\left[  e^{i\pi k_{1}\cdot k_{2}} A_{L}(2, 1, 3, 4) + A_{L}( 1, 2, 3, 4)   +   e^{i\pi k_{2}\cdot k_{3}}A_{L}(1, 3, 2,  4) \right]\nonumber\\
    &A_{R}(1, 3, 2, 4)\left[  e^{i\pi  k_{2}\cdot (k_{1} + k_{3})} A_{L}(2, 1, 3, 4) +e^{i\pi k_{2}\cdot k_{3}} A_{L}( 1, 2, 3, 4)  + A_{L}(1, 3, 2,  4) \right].\nonumber\\
\end{align}

The subscript $L$ in $A_{L}$ indicates that the sign of $\alpha{}'$ is not changed, while $R$ in $A_{R}$ indicates that the sign of $\alpha{}'$ is changed. 

The expression above could be greatly simplified by using the monodromy relations of open string amplitudes \cite{Stieberger:2009hq,Bjerrum-Bohr:2009ulz}. It could be derived from performing certain closed contour integrals in the open string amplitude. It reads,
\begin{equation}
    A(1, 2, \cdots , n ) + e^{i\pi k_{1}\cdot k_{2}}A(2, 1, \cdots, n) + \cdots  e^{i\pi k_{1}\cdot (k_{2}+\cdots k_{n-1})}A(2, \cdots, 1, n) = 0.
\end{equation}
In particular, the first line and third line of (\ref{4ptraw}) will vanish due to the monodromy relations. On the other hand, the second line would also greatly simplify. We get (after writing all the momenta in terms of Mandelstam variables)
\begin{equation}\label{4ptamp}
    M(s, t,  u) = \sin(\pi s)A(s, u)A(-s, -t).
\end{equation}

This expression doesn't look like $s, t, u$ permutation invariant, one could verify that it indeed is by using the monodromy relations again. (It is written a bit differently).
\begin{equation}
       A(s, t) = \frac{\sin(\pi u)}{\sin(\pi t)}A(s, u), \quad A(u, t) = \frac{\sin(\pi s)}{\sin(\pi t)}A(s, u).
\end{equation}

The explicit form of the amplitude could also be calculated. The following expression for 4-point open string (Yang-Mills) amplitude is the most convenient for us. 
\begin{equation}
    A(s, t) = K\frac{\Gamma(-s)\Gamma(-t)}{\Gamma(1-s-t)}.
\end{equation}
The kinematic factor $K$ is a function of the polarization vectors, and its exact form changes depend whether we are considering superstring theory or bosonic string theory, which is given in \cite{Feng:2004tg}.
From the properties of the gamma function, we could see that $A(s, t)$ have poles at $s = n$ where $n$ is a non-negative integer. These correspond to the higher spin resonance states of string theory.
Therefore for the 4-point chiral string amplitude (\ref{4ptamp}), all poles in $s$ are canceled by $\sin(\pi s)$  except for $s = 0$, therefore no-higher spin resonance state exists.\footnote{For bosonic strings, the kinematic factor $K$ have poles at $s = -1$, which correspond to tachyon exchanges. In this case, the chiral string amplitude would also contain massive states and tachyon states. }

Using the gamma function identity $\Gamma(z)\Gamma(1-z) = \pi/(\sin\pi z)$, the chiral string amplitude become
\begin{equation}
    M(s, t, u) = K_{R}K_{L}\frac{1}{stu}.
\end{equation}
Where $K_{L}$ is just the kinematic factor $k$ but $K_{R}$ is the same kinematic factor but $\alpha{}'$ is flipped. this is indeed the graviton 4-point amplitude.\cite{Sannan:1986tz}.

\subsection{Spectrum}
\label{spectrum}
We would like to emphasize that the fact we get field theory amplitude from changing the string propagator is not a coincidence. The chiral string theory indeed only contains a finite number of states. We could see that by analyzing its spectrum. This could be done by finding the vertex operators with correct holomorphic dimensions. This is similar in spirit to the usual bosonic/superstring theory.

We shall only consider a bosonic theory for brevity. The most general vertex operator that one could construct is the following form
\begin{equation}
    V^{(N, M)}(z, \bar{z}; k) = \epsilon_{\mu_{1}\cdots \mu_{n}}{}^{\nu_{1}\cdots \nu_{m}}\prod^{n}_{j}\partial^{n_{j}} X^{\mu_{j}} \prod^{m}_{i}\bar{\partial}^{m_{i}} X_{\nu_{i}} e^{i k \cdot X}.
\end{equation}
Non-negative integers $(N, M)$ label the mass level of such vertex operators, where $N = \sum_{j} n_{j},  M = \sum_{i} m_{i}$. Note that the on shell condition for the momentum $k$ is that $V^{(N, M)}$ has holomorphic dimensions $(1, 1)$. The holomorphic dimensions could be calculated from the vertex operators OPE with energy-momentum tensor $T(z), \overline{T}(\overline{z})$ (see \cite{Li:2022tbz} for stress tensor OPE). 
\begin{align*}
    &T(z)V(0) \sim \frac{\partial T(0)}{z} +\frac{hT(0)}{z^{2}}\nonumber\\
    &\overline{T}(\overline{z})V(0) \sim \frac{\overline{\partial} T(0)}{\overline{z}} +\frac{\overline{h}T(0)}{\overline{z}^{2}}.
\end{align*}

For usual string theory, we have $(h, \bar{h}) = (N + k^{2}, M + k^{2}) = (1, 1)$. This gives us the mass shell and level matching condition $N = M$, $k^{2} = 1-N$. For chiral strings, the $XX$ OPE is modified, thus the holomorphic dimensions of vertex operators will also be modified. Given the same vertex operator $V^{(N, M)}$ as above, its holomorphic dimensions becomes
\begin{equation}
    h = N + k^{2},\quad \bar{h} = M - k^{2}.
\end{equation}
Demanding that $(h, \bar{h}) = (1, 1)$ fixes three possible mass level. $(N, M) = (1, 1), (0, 2)$ and $(2, 0)$. They correspond to one massless sector, one massive sector, and one tachyonic sector (see \cite{Lee:2017utr} for the norm of these states). 
\begin{center}
\begin{tabular}{|c| c| c| c|} 
 \hline
 $N$ & $M$ & $m^2$ & state \\ [0.5ex] 
 \hline
 1 & 1 & 0 & $\varepsilon_{\mu\nu}\alpha^{\mu}_{-1}\bar{\alpha}^{\nu}_{1}|0,k\rangle$ \\ 
 \hline
 2 & 0 & $\frac{1}{\alpha'}$ & $a_{\mu\nu}\alpha^{\mu}_{-1}\alpha^{\nu}_{-1}|0,k\rangle$ \\
 \hline
 0 & 2 & $-\frac{1}{\alpha'}$ &$\bar{a}_{\mu\nu}\bar{\alpha}^{\mu}_{-1}\bar{\alpha}^{\nu}_{-1}|0,k\rangle$ \\
 \hline
\end{tabular}
\end{center}

The same kind of analysis could also be done for superstring theories. For type II theories where the ground states are massless, the respective condition is $(h,\bar{h})= (N+k^2,M+k^2)=(0,0)$. For the chiral type II theories these conditions will get modified to $(h,\bar{h})= (N+k^2,M-k^2)=(0,0)$. These conditions fix the state to be $(N,M)=(0,0)$ which is massless. This state can be decomposed into graviton, Kalb-Ramond, and dilaton fields. One can construct the vertex operators corresponding to these states on the world-sheet. 

\subsection{HSZ gauge}
\label{hsz gauge}
If we take the chiral strings to the  Hohm-Siegel-Zwiebach \cite{Hohm:2013jaa} (HSZ) gauge, then the calculations, such as the OPE of vertex operators become easier.

Consider the following general coordinate transformation on the world-sheet
\begin{eqnarray}
\bar{z}\rightarrow\sqrt{1+\beta}\bar{z},\,\,\,\,
 z\rightarrow  \frac{1}{\sqrt{1+\beta}}(z-\beta \bar{z}),
\end{eqnarray}
$\beta$ is an arbitrary real parameter. $\beta = 0$ is just the usual conformal gauge, and if we take $\beta \rightarrow \infty$, it becomes the HSZ gauge. Under this gauge, the string action becomes.

\begin{eqnarray}
\mathcal{L}=-\frac{1}{2}[\beta(\bar\partial X)(\bar\partial X)+(\bar\partial X)(\partial X)]
\end{eqnarray}

The chiral string propagator also changes under the coordinate transformation. After discarding an irrelevant constant (By momentum conservation, the constants in propagators don't contribute to the ampliutde.) and expanding the propagator in $1/\beta$, it becomes
\begin{eqnarray}
    G(z_{+},z_{-})=\ln\Bigg(\frac{z_{+}}{z_{-}-i\epsilon/z_{+}}\Bigg) \rightarrow \ln\Bigg[ \frac{z_{+} -\beta z_{-}}{z_{-} -i\epsilon/(z_{+} - \beta z_{-})} \Bigg] \xrightarrow{\beta \rightarrow \infty} -\frac{z_{+}}{\beta z_{-}}
\end{eqnarray}
As usual, we have performed the analytic continuation (\ref{acont}).
While the propagator seems to vanish in the $\beta \rightarrow \infty$ limit,  we will see that the amplitude from this propagator won't. As $\beta$ will serve as an IR regulator for $z_{+}$ integration. One observation which is relevant is $z_{+}$ always comes with $\frac{z_{+}}{\beta}$.\\

The propagators of ghosts, fermions, super-ghosts, etc, would also need to be modified according to the new coordinate transformations. These modifications can be summarized by following replacement rules for the $1/z$ and $1/\bar{z}$ propagators in (\ref{ghostprop}) and (\ref{fermionprop}).  After truncating the expression at first order in $\frac{z_{+}}{\beta}$
\begin{eqnarray}
   \frac{1}{z_{-}}\rightarrow\frac{1}{z_{-}}, \,\,\,\,\,
\frac{1}{z_{+}}\rightarrow \frac{1}{\beta}\Bigg(\frac{1}{z_{-}}+\frac{z_{+}}{\beta z_{-}^2}\Bigg)
\label{eq:def} 
\end{eqnarray}
 Let's consider n-point tachyon amplitude as a primitive example.
\begin{eqnarray}
    \prod_{i=1}^n\langle e^{i k_i.X(z_i)}\rangle = e^{S_0}, \quad S_0=\frac{1}{2 \beta} \sum_{i,j}k_i.k_j \frac{z_{ij+}}{z_{ij-}}=\frac{1}{\beta}\sum_i z_{i+} \sum_j \frac{k_i.k_j}{z_{ij-}}
\end{eqnarray}
Now, the world-sheet integral over $z_{i+}$ become (using a further analytic continuation from $z_{i+}$ to $iz_{i+}$)
\begin{eqnarray}
    \int_{-\infty}^{\infty} d^{n-3}z_{i+} e^{S_0}\sim \beta^{n-3} \prod_i \delta^d\Bigg(\sum_j \frac{k_i.k_j}{z_{ij-}}\Bigg)
\end{eqnarray}

These delta functions are just factors that will enforce the scattering equations, just like in the CHY formula and the ambitwistor strings. To get the correct amplitude, one needs to carefully prescribe the integrated vertex operators. Fortunately in our paper, we study the celestial holography directly in the OPEs of the vertex operators, so we could just use the usual vertex operators directly \footnote{The amplitude could always be written as $M_{n} = \mathcal{V}_{\textrm{CKG}}^{-1}\int \langle V_{1}\cdots V_{n }\rangle$ where $V_{i}$ is just the bare vertex operators without any ghost. Of course, if we want to calculate the amplitude, this expression wouldn't work as it will give indeterminate $\infty/\infty$. That's why we need to specify the BRST formalism to get a gauge independent amplitude. But in doing everything in terms of vertex operators, the OPEs are not sensitive to these divergences.  }, bypassing the need to calculate the amplitude. Therefore, the construction of the amplitude in the HSZ gauge will not be reviewed here. For details, see \cite{LipinskiJusinskas:2019cej}.

\section{Energetic soft theorem from Chiral string world-sheet}
\label{energetic soft theorem}
The soft theorem is a statement about amplitudes with $n+1$ particles with $n$ particles having finite momenta and one particle having soft momenta $k$ with energy $\omega \rightarrow 0$ and amplitudes with $n$ particles times the soft factor.
\begin{eqnarray}
   \lim_{\omega\rightarrow 0} \mathcal{A}_{n+1}(p_1 \cdots p_n, k) = S(p_i,k)  \mathcal{A}_{n}(p_1 \cdots p_n)
\end{eqnarray}
The leading and subleading soft theorems are universal for any perturbative quantum field theories coupled with gravity  \cite{Sen:2017nim, Laddha:2017ygw,Krishna:2023fxg}. \footnote{Generally this requires the space-time dimension to be greater than 4.}  But in non-abelian gauge theory where even leading soft theorem gets modified due to loop effects \cite{Bern:2014oka}.
The loop corrections will modify the soft expansion in terms of $\omega^i \mathrm{Log}^j[\omega]$ \cite{Ghosh:2021bam,Sahoo:2020ryf,Sahoo:2021ctw}. But In this paper, we will look at the tree-level soft theorems as we will do the OPE on the sphere. 

In this section, we will re-derive the soft theorems for bosonic strings and (super)-chiral strings amplitude. We will discover that the leading and sub-leading factors could be completely determined by the OPEs of vertex operators. Although the OPEs of chiral strings and usual strings are in general different, we will see that the leading and subleading soft theorem doesn't depend on the usual string/chiral string theory. This gives us an alternative explanation of how the soft theorems could arise in quantum field theories that have a world-sheet description, and lead us to a natural definition of ``soft vertex operators.''

 We will study the soft theorems bosonic soft chiral strings in detail in section \ref{sec:3-1}, calculate the soft factors and present the soft vertex operators for the super chiral strings in HSZ gauge in section \ref{soft theorem in superstring hsz}. \\

\subsection{Soft limit form chiral string OPEs}
\label{sec:3-1}
One can compute the soft theorems directly from string world-sheet \cite{DiVecchia:2015oba,Higuchi:2018vyu,Huang:2016bdd} without going to the amplitude formulation. We will proceed similarly in the chiral case.
The scattering amplitudes in a closed string case on the sphere are
\begin{eqnarray}
\mathcal{M}(1,2,...,n)=\int_{S^2} dz_2...dz_{n-2}\langle c\tilde{c}V(1)V(2)...c \tilde{c}V(n-1)c \tilde{c} V(n)\rangle
\end{eqnarray}
Here $V$ is the closed string vertex operator and $c$ is the ghost. For example, the graviton vertex operator can be written as
\begin{equation}
    V_{G}(z, \bar{z}; q) = e_{\mu\nu}\partial X^{\mu} \bar{\partial} X^{\nu} e^{i q \cdot X(z, \bar{z})}.
\end{equation}

Since the vertex operators are integrated over the world-sheet, we can use partial integration to rewrite the vertex operators as
\begin{equation}
    V_{G} = \frac{1}{2}e_{\mu\nu}X^{\mu}X^{\nu}\partial\bar{\partial}e^{iq\cdot X}.
\end{equation}

In the soft limit when $q \rightarrow 0$, we have $\partial\bar{\partial}e^{iq\cdot X} = iq\cdot \partial\bar{\partial}X + O(q^{2})$. Therefore, in the soft limit, the $O(q)$ terms are proportional to the equation of motion and therefore vanish unless the soft vertex operators collide with other vertex operators, where the OPE expansion is applicable. This is why the OPEs alone can determine the soft theorems.\\

To be more specific, we will divide the integration of the soft vertex operators into regions that are near the other vertex operators and regions that are far away from other vertex operators.
\begin{eqnarray}\label{vdcomp}
    \int_{\mathbb{C}}d^2zV_{s}(z, \bar{z}; q) = \sum_{i}^{n}\int_{\mathcal{B}_{I}}d^2zV_{s}(z, \bar{z}; q) + \int_{\mathbb{C} / \cup \mathcal{B}_{i}}d^2zV_{s}(z, \bar{z}; q).
\end{eqnarray}

Where $\mathcal{B}_{i} \equiv \{z| : |z-z_{i}| < \epsilon \}$, and $z_{i}$ are positions of $i$-th pther vertex operators. From the discussion above, up to sub-subleading soft terms. we only need to consider the former terms (regions $\mathcal{B}_{I}$) in (\ref{vdcomp}), and that's why it's possible to factor out the soft factors up to the sub-subleading order.  We shall take the $\epsilon \rightarrow 0$ limit after we calculate the OPEs. 

A generic vertex operator in bosonic theories could be written as 
\begin{equation}
    V_{h}^{(s)}(z, \bar{z}; k) = A_{\mu_{1}\cdots \mu_{n}}^{(s)\hspace{15pt}\nu_{1}\cdots \nu_{m}}(k)\prod^{n}_{j}\partial^{n_{j}} X^{\mu_{j}} \prod^{m}_{i}\bar{\partial}^{m_{i}} X_{\nu_{i}} e^{i k \cdot X}
\end{equation}

We would like to find the OPE of this hard operator with a graviton soft operator. As usual, we could use the exponentiation trick.
\begin{equation}
    V_{h} \rightarrow \overline{V}_{h} \equiv \exp i\left(  k \cdot X + \sum_{j}^{n}\xi_{j}\cdot\partial^{n_{j}}X  +  
   \sum_{i}^{m}\lambda_{i}\cdot\bar{\partial}^{m_{i}}X \right),
\end{equation}

The exponentiation for $V_{s} = X^{\mu}X^{\nu}\partial\bar{\partial}e^{iq\cdot X}$ is a little more tricky, we introduce an extra coordinate $z{}'$ and a dummy polarization vector $\zeta^{\mu}$. 
\begin{equation}
    V_{s} \rightarrow \overline{V}_{s}(z, z^{'}) \equiv \exp \left(i q \cdot X(z) + i\zeta \cdot X(z^{'})\right). 
\end{equation} 
After we evaluate the OPE of $\overline{V}_{s}$ with other operators, we need to act $\partial_{z}\bar{\partial}_{z}$ on it and extract the quadratic components in $\zeta$ to get the desired result.

The contraction between $V_{h}$ and $V_{s}$ can be obtained from the contraction between $\overline{V}_{h}$ and $\overline{V}_{s}$.
\begin{equation}
\contract{V_{s}(z)}\contract{V_{h}(w)}  = -\frac{1}{2}e_{\mu\nu} \lim_{\zeta \rightarrow 0} \partial_{\zeta}^{\mu} \partial_{\zeta}^{\nu} \lim_{z \rightarrow z^{'}} \partial_{z}\partial_{\bar{z}}\left.\contract{\overline{V}_{s}(z, z^{'})}\contract{\overline{V}_{h}(w)}\right|_{\textrm{ multi-linear in $i\xi_{i}$ and $i\lambda_{i}$}}.
\end{equation}
The explicit calculation yields
\begin{align}
    &\contract{\bar{V}_{s}(z, z^{\prime}) } \contract{\bar{V}_{h}(w)} = \\
    & \; (z - w)^{\frac{1}{2} k \cdot q}(z^{\prime} - w)^{\frac{1}{2}k \cdot \zeta }(\bar{z} - \bar{w})^{\frac{\lambda}{2} k \cdot q}(\bar{z}^{\prime} - \bar{w})^{\frac{1}{2} k \cdot \zeta } \nonumber \\
    &\exp \left[-\frac{1}{2}\sum_{i}^{n}\frac{(n_{i} -1)! q \cdot \xi_{i}}{(z-w)^{n_{i}}} -\frac{\lambda}{2}\sum_{i}^{m}\frac{(m_{i} -1)! q \cdot \lambda_{i}}{(\bar{z}-\bar{w})^{m_{i}}}\right]\nonumber\\
    &\exp \left[-\frac{1}{2}\sum_{i}^{n}\frac{(n_{i} -1)! \zeta \cdot \xi_{i}}{(z^{\prime}-w)^{n_{i}}} -\frac{\lambda}{2}\sum_{i}^{m}\frac{(m_{i} -1)! \zeta \cdot \lambda_{i}}{(\bar{z}^{\prime}-\bar{w})^{m_{i}}}\right]\nonumber\\
    &\contract{\exp i\left[ q \cdot X(z) + \zeta \cdot X(z^{'}) +  k \cdot X(w) + \sum_{j}^{n}\xi_{j}\cdot\partial^{n_{j}}X(w)  + \sum_{i}^{m}\lambda_{i}\cdot\bar{\partial}^{m_{i}}X(w)\right]}\nonumber 
\end{align}
In the above expression, the $\lambda  = 1$ corresponds to the string theory case while $\lambda = -1$ corresponds to the chiral string case. Apply partial derivatives $\partial \bar{\partial}$ on the results and take the $z^{\prime} \rightarrow z$ limit, we get
\begin{align*}
 & \left.
    \begin{aligned}
        &(z - w)^{\frac{1}{2}k\cdot (q + \zeta)}(\bar{z} - \bar{w})^{\frac{\lambda}{2}k\cdot (q + \zeta)}\\
        &\exp \left[-\frac{1}{2}\sum_{i}^{n}\frac{(n_{i} -1)! (q + \zeta) \cdot \xi_{i}}{(z-w)^{n_{i}}} -\frac{\lambda}{2}\sum_{i}^{m}\frac{(m_{i} -1)! (q + \zeta) \cdot \lambda_{i}}{(\bar{z}-\bar{w})^{m_{i}}}\right]\\
        &\contract{\exp i\left[ (q + \zeta) \cdot X(z)  +  k \cdot X(w) + \sum_{j}^{n}\xi_{j}\cdot\partial^{n_{j}}X(w)  + \sum_{i}^{m}\lambda_{i}\cdot\bar{\partial}^{m_{i}}X(w)\right]}
    \end{aligned}
\right\} \quad \textrm{(I)}\\
 & \left.
    \begin{aligned}
        &\left[\frac{k\cdot q}{2(z - w)} + \sum_{i}^{n}\frac{n_{i}!q\cdot \xi_{i}}{(z - w)^{n_{i}}} + iq \cdot \partial X(z)\right]\\
        &\left[\frac{\lambda k\cdot q}{2(\bar{z} - \bar{w})} + \sum_{i}^{m}\frac{\lambda m_{i}!q\cdot \lambda_{i}}{(\bar{z} - \bar{w})^{m_{i}}} + iq \cdot \bar{\partial} X(\bar{z})\right]
    \end{aligned}
\right\} \quad \textrm{(II)}
\end{align*}

To get the OPE, one needs to expand the above expression in powers of $(z-w)$ and then extract the multi-linear parts in $i\xi_{i}$ and $i\lambda_{i}$ and the quadratic part of $\zeta$. Before we get into the explicit calculation, a few observations make the calculation easier.

The expression above can be split into two parts, where we label them as (I) and (II). (I) only depends on $q+\zeta$, so we could first expand it in terms of $q + \zeta$, then expand in $q$. On the other hand, (II) is proportional to $q^{2}$. That means if we want to find up to sub-subleading soft theorems (which are only proportional to $q$), we only need to consider terms singular in $q$ inside (I) when doing the expansion. More explicitly,  the expansion of (I) and (II) have the following form
\begin{equation}
\begin{split}
    \textrm{(I)}=&\sum_{a_{i}, b_{i}}C_{a_{i}, b_{i}}(q+\zeta, \xi_{i}, \lambda_{i})(z - w)^{\frac{1}{2}k\cdot (q + \zeta) + a_{i}}(\bar{z} - \bar{w})^{\frac{\lambda}{2}k\cdot (q + \zeta) + b_{i}}\\
    \textrm{(II)}=& \sum_{a_{i}, b_{i}} D_{a_{i}, b_{i}}(q, \xi_{i}, \lambda_{i})(z - w)^{a_{i}}(\bar{z} - \bar{w})^{b_{i}}
\end{split}
\end{equation}

$D_{a_{i}, b_{i}}$ are proprotional to $q^{2}$, and that $C_{a_{i}, b_{i}}$ are not singular in $q$, that means the only way for terms singular in $q$ appear in (I) is after the integration of $z$ in ($\ref{vdcomp}$).

To treat chiral strings and usual strings uniformly, we shall perform the integration differently from \cite{Higuchi:2018vyu}; We do a Wick rotation on the world-sheet coordinate, therefore $z, \bar{z} \in \mathbb{C}\;\Rightarrow\; z^{+}, z^{-} \in \mathbb{R}$

The integration over the world-sheet becomes integration over contours of $z^{\pm}$.
\begin{align}
    & \int d^{2}z_2...d^{2}z_{n-2}\enspace\langle c\tilde{c}V(1)V(2)...c \tilde{c}V(n-1)c \tilde{c} V(n)\rangle\\
    \Rightarrow& \sum_{\sigma \in \textrm{S}_{n-3}} \int_{0 \leq z^{-}_{\sigma(2)}  \cdots \leq z^{-}_{\sigma(n-2)} \leq 1}  dz^{-}_{\sigma(2)} \cdots  dz^{-}_{\sigma(n-2)} \enspace \int_{C_{2}} dz^{+}_{\sigma(2)} \cdots \int_{C_{n-2}} dz^{+}_{\sigma(n-2)}\nonumber\\
    &\langle\; c\tilde{c}V(0)V(z_{2}^{+}, z_{2}^{-})\;\cdots \;V(z_{n-2}^{+}, z_{n-2}^{-})c \tilde{c}V(1)c \tilde{c} V(\infty)\;\rangle
\end{align}
This step is in fact the first step to derive the KLT relations \cite{Kawai:1985xq} , and the contours $C_{2}\cdots C_{n-2}$ are given in figure \ref{fig:1} \cite{Bjerrum-Bohr:2010pnr}. 
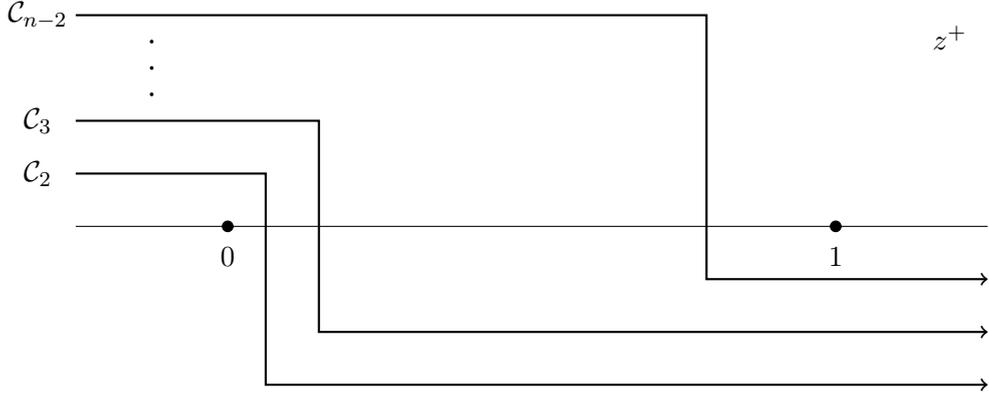
\begin{figure}[h]
    \centering
\begin{tikzpicture}
    \draw (-6,0) -- (6,0);
    \draw[thick, ->] (-6, 0.7) -- (-3.5, 0.7 ) -- (-3.5, -2.1) -- (6, -2.1);
    \draw[thick, ->] (-6, 1.4) -- (-2.8, 1.4 ) -- (-2.8, -1.4) -- (6, -1.4);
    \draw[thick, ->] (-6, 2.8) -- (2.3, 2.8 ) -- (2.3, -0.7) -- (6, -0.7);
    \filldraw [black] (-4,0) circle (2pt);
    \filldraw [black] (4,0) circle (2pt);
    \filldraw [black] (-5, 1.75) circle (0.5pt);
    \filldraw [black] (-5, 2.1) circle (0.5pt);
    \filldraw [black] (-5, 2.45) circle (0.5pt);
    \draw  (-4,-0.4) node {0};
    \draw  (4,-0.4) node {1};
    \draw  (-6.5, 0.7) node {$\mathcal{C}_{2}$};
    \draw  (-6.5, 1.4) node {$\mathcal{C}_{3}$};
    \draw  (-6.5, 2.8) node {$\mathcal{C}_{n-2}$};
    \draw  (5.5, 2.5) node {$z^{+}$};
\end{tikzpicture}

\caption{\label{fig:1} The shape of contours  $\mathcal{C}_{2}, \cdots , \mathcal{C}_{n - 2}$ when $ 0 \leq z^{-}_{2}  \cdots \leq z^{-}_{n-2} \leq 1 $.  To get the full amplitude, we need to sum over all possible permutations of $z^{-}$s, with all contours $\mathcal{C}_{i}$ permutes accordingly. }
\end{figure}

Without loss of generality, we assume that the second vertex operator $V(z_{2}^{+}, z_{2}^{+})$ becomes soft, and if we fixed the positions of all other vertex operators and only focus on integration over $C_{2}$. A typical shape of the contour looks like figure \ref{fig:2}.

Note that we want to only want to integrate over the region where the soft vertex operators collide with another vertex operator. If we pull the contour to the left, then one could see that we only need to only consider segments of the contour where it is in the vicinity of the vertex positions to the left (see figure \ref{fig:2} again). What is not shown in this figure is that we also need to integrate the $z^{-}$ line where $z^{-}_{2}$ is close to $z^{-}_{i}$ and since there is no other vertex between $z^{-}_{2}$ and $z^{-}_{i}$, $z^{+}_{i}$ should be right next to $C_{2}$ contour to the left, and the part of the contour in between the two dashed line is the only region we need to consider 
when $V_{2}$ and $V_{i}$ collide.

\begin{figure}[h]
    \centering
    \begin{tikzpicture}
        \draw (-7,0) -- (-1,0);
        \filldraw [black] (-6,0) circle (2pt);
        \filldraw [black] (-2,0) circle (2pt);
        \draw[thick, ->] (-7, 0.5) -- (-3.05, 0.5 ) -- (-3.05, -0.5) -- (-1, -0.5);
        \draw (-3.75, 0) node[cross] {};
        \draw (-2.5, 0) node[cross] {};
        \draw (-5.5, 0) node[cross] {};

        \draw (1,0) -- (7,0);
        \filldraw [black] (2,0) circle (2pt);
        \filldraw [black] (6,0) circle (2pt);
        \draw[thick] (1, 0.5) -- (4.75, 0.5 );
        \draw[thick] (4.75, 0.5) arc (90:-90:0.5);
        \draw[thick, <-] (1, -0.5) -- (4.75, -0.5);
        \draw (4.75, 0) node[cross] {};
        \draw (5.5, 0) node[cross] {};
        \draw (2.5, 0) node[cross] {};
        \draw[dashed] (4.25, -1) -- (4.25, 2);
        \draw[dashed] (5.25, -1) -- (5.25, 2);
        \draw[<->, thick] (4.25, 1.5) -- (5.25, 1.5);

        \draw  (4.75, -0.3) node {$z^{+}_{i}$};
        \draw  (-3.75, -0.3) node {$z^{+}_{i}$};

        \draw  (-4, 1) node {$\mathcal{C}_{2}$};
        \draw  (4.75, 1.8) node {$\varepsilon$};
        \draw  (0, 0) node {$\Longrightarrow$};
       
    \end{tikzpicture}
   \caption{\label{fig:2} The shape of soft contour $\mathcal{C}_{2}$, the region between two dashed lines and $z^{-}_{i} < z^{-}_{2} < z^{-}_{i} + \epsilon$ is the region where the soft vertex operators collides with $V(z_{i})$. Note that $z_{2}^{-} < z^{-}_{i}$ dose not contribute since in this region, $\mathcal{C}_{2}$ cannot be close $z_{i}^{+}$. }
\end{figure}
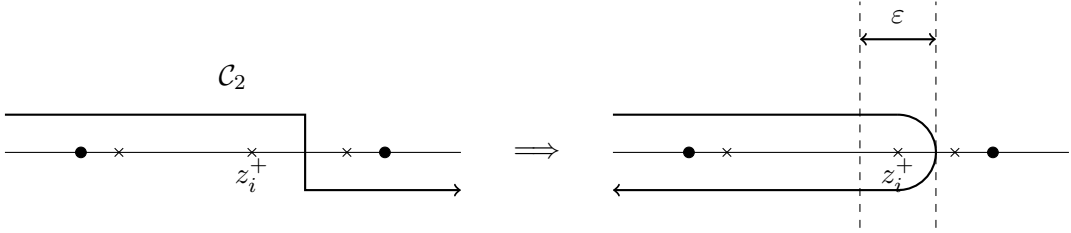

Suppose that after performing the OPE, we have a term that is proportional to 
\begin{equation}\label{opepwer}
(z^{+} - w^{+})^{\frac{1}{2}k\cdot (q + \zeta) + n}(z^{-} - w^{-})^{\frac{\lambda}{2}k\cdot (q + \zeta) + m}.
\end{equation}
We need to integrate this over the region we just specified above; the result is
\begin{align*}
    &2i\sin\pi k \cdot (q + \zeta) \int_{w^{-}}^{w^{-} + \epsilon} dw^{-} |z^{-} - w^{-}|^{-\frac{1}{2}k \cdot (q + \zeta) + n} \int_{w^{+}}^{w^{+} + \epsilon} dw^{+} |z^{+} - w^{+}|^{-\frac{\lambda}{2}k \cdot (q + \zeta) + m}\\
    =&2i\left[\pi k\cdot(q + \zeta) + O((q + \zeta)^{3}) \right] \frac{\epsilon^{-\frac{1}{2}k \cdot (q + \zeta) + n + 1} }{(-\frac{1}{2}k \cdot (q + \zeta) + n + 1)} \frac{\epsilon^{-\frac{\lambda}{2}k \cdot (q + \zeta) + m + 1} }{(-\frac{\lambda}{2}k \cdot (q + \zeta) + m + 1)}
\end{align*}

We can see that only when $m = -1, n = -1$, the expression is singular in $q$, and this contribution is 
\begin{equation}\label{sigterm}
8i\pi \frac{\epsilon^{\frac{(1+\lambda)}{2} k\cdot (q + \zeta)}}{\lambda k\cdot(q + \zeta)}.
\end{equation}
\emph{Therefore, in the soft limit, we only need to include terms of OPE expansions where $m, n = -1$ in (\ref{opepwer})}.
\footnote{A more abstract way to see that only terms with $n, m = -1$ could contribute if we perform the OPE between a soft vertex operator and a hard vertex operator. Only primaries that are in the BRST cohomology could survive inside correlators. The condition for the vertex operators to be in the BRST cohomology coincides with the on-shell condition this further implies that the vertex operators must have holomorphic dimensions $(1, 1)$. This means that the powers of the OPE expansion are $h_{k} - h_{i} -h_{j} = -1$. (The $k \cdot (q + \zeta)$ term vanish because we are taking the soft limit)}

If we expand (\ref{sigterm}) in the powers of $\zeta$.
\begin{equation}\label{sigexpand}
    \frac{1}{k\cdot q}\left(1 - \frac{\zeta \cdot k}{k \cdot q} + \left(\frac{\zeta \cdot k}{k \cdot q}\right)^{2} + \cdots\right)
\end{equation}

Since we only need to extract quadratic terms in $\zeta$, terms beyond $O((k\cdot q)^{-3})$  can be ignored. Also, the dependence of $\zeta$ only comes from (I) as a function of $q + \zeta$. Using this fact, the calculation of expansions could be done in a systematic manner that simplifies the calculation.

\begin{itemize}
    \item If we want the leading soft theorem, we only need to expand to the zeroth degree of $q+\zeta$ in (I) and only include the 3rd term of (\ref{sigexpand}).
    \item  If we want to have the sub-leading soft theorem, we need to expand up to 1st degree of $q+\zeta$ in (I), and include 2nd and 3rd terms in (\ref{sigexpand}).
    \item  Finally if we want sub-subleading terms, we need to expand up to 2nd degree in $q+\zeta$ in (I), and include the 1st, 2nd, and 3rd terms in (\ref{sigexpand}).
\end{itemize}.

The rest of the calculations are straightforward but tedious, and they are exactly the same as in \cite{Kawai:1985xq}, so we won't elaborate on the calculations here. We find the standard leading and subleading soft theorem, thus proving the universality of the soft theorem for string and chiral strings.

$$
\int d^{2}z \contract{V_{s}(z; q) } \contract{V_{h}(w; k)} \sim [ S^{(0)} + S^{(1)}] \contract{V_{h}(w; k)},
$$
where
\begin{align*}
    S^{(0)} & = \frac{e_{\mu\nu} p^{\mu}p^{\nu}}{ p \cdot q}, \quad S^{(1)} 
     = \frac{e_{\mu\nu} p^{\mu}q_{\rho} J^{\rho \nu}}{ p \cdot q} \\
    J^{\mu\nu} & = L^{\mu\nu} + S^{\mu\nu}, \quad L^{\mu\nu}  = p_{\mu}\frac{\partial}{p^{\nu}} - p_{\nu}\frac{\partial}{p^{\mu}}.
\end{align*}
and $S_{\mu\nu}$ is the spin operators that acted on $V_{h}$ according to its spin. 
\subsection{Soft limit for super chiral strings in HSZ gauge.}
\label{soft theorem in superstring hsz}
In this section, we derive the soft theorems for super-chiral strings theories in the HSZ gauge. We shall derive them directly from the OPEs of vertex operators. In superstring theory we have fermions and doing the OPE by the above method will become very complicated (See section 4 of \cite{Jiang:2021csc} nevertheless) Therefore, we will take the HSZ gauge limit instead. Although there is no explicit proof that taking this gauge limit will result in the same theories, from  \cite{LipinskiJusinskas:2019cej} we know that chiral strings in HSZ gauge reproduce QFTs with a finite number of particles. 

In the superstring theory, In order to have a non-vanishing amplitude, one also needs to saturate the super-ghost charge to $-2$. This is also true in chiral string. We will take the soft operator in the $q = 0$ picture and the two hard operators in the $q = -1$ picture. Here, We will only do the OPE of gravitons. The OPE of other operators can be done but is cumbersome.\\

The soft graviton vertex operator in $(0,0)$ picture and hard graviton vertex operator in $(-1,-1)$ picture can be written as \cite{Polchinski:1998rr}
\begin{eqnarray}
    V_s^{(0,0)}(z,\bar{z})&\equiv&e_{\mu\nu}(i \partial X^{\mu}+k_s\cdot\psi \psi^{\mu})(i \bar{\partial}X^{\nu}+k_s\cdot\tilde{\psi}\tilde{\psi}^{\nu})e^{i k_s \cdot X}\nonumber\\
V_h^{(-1,-1)}(w,\bar{w})&\equiv&e_{\mu\nu}\psi^{\mu} \tilde{\psi}^{\nu} e^{-\varphi}e^{-\tilde{\varphi}}e^{i k_i \cdot X}
\end{eqnarray}

here $e_{\mu\nu}=\varepsilon_{\mu}\tilde{\varepsilon}_{\nu}$ is the polarization of the gravitons. We are interested in the soft limit ($k_s \rightarrow 0$). In this limit the vertex operator $V_{s}^{(0,0)}(z)$ is a total derivative as in bosonic case. This can be seen upon expanding the vertex operators
\begin{eqnarray}
    V_s^{(0,0)}(z,\bar{z})&&= e_{\mu\nu}\Bigg( i \partial X^{\mu} i \bar{\partial} X^{\nu} e^{i k_s \cdot X}+ i \partial X^{\mu} k_s\cdot\tilde{\psi}\tilde{\psi}^{\nu}e^{i k_s \cdot X}+  i \bar{\partial} X^{\nu} k_s\cdot\psi \psi^{\mu} e^{i k_s \cdot X} +\mathcal{O}(k_s^2)\Bigg)\nonumber\\
  &&=  e_{\mu\nu}\Bigg( i  X^{\mu} i  X^{\nu} \partial\bar{\partial} e^{i k_s \cdot X}- i X^{\mu} \partial (k_s \cdot \tilde{\psi} \tilde{\psi}^{\nu}e^{i k_s \cdot X})-i X^{\nu} \bar{\partial} (k_s \cdot \psi \psi^{\nu}e^{i k_s \cdot X})+\mathcal{O}(k_s^2)\Bigg)\nonumber\\
  &&= e_{\mu\nu}\Bigg( i^2 X^{\mu}X^{\nu} \partial\bar{\partial} (i k_s \cdot X)-i X^{\mu} \partial (k_s \cdot \tilde{\psi} \tilde{\psi}^{\nu})-i X^{\nu} \bar{\partial} (k_s \cdot \psi \psi^{\nu})+\mathcal{O}(k_s^2)\Bigg)\nonumber\\
\end{eqnarray}
All these terms are proportional to the equation of motion hence the OPE of this operator with hard vertex operators will be non-vanishing only when soft operators collide with hard ones.
The OPE of soft $V_{s}^{(0,0)}(z,\bar{z})$ with hard $V_{h}^{(-1,-1)}(w,\bar{w})$ in the chiral superstring theory can be written as (following from the slight modification of the superstring case in \cite{Bianchi:2014gla}.)

\begin{eqnarray}
V_{s}^{(0,0)}(z_s,\bar{z}_s)V_{h}^{(-1,-1)}(z_i,\bar{z}_i)&\sim& (z_s-z_i)^{q\cdot k}(\bar{z}_s-\bar{z}_i)^{  -q\cdot k}(z_s-z_i)^{-1}(\bar{z}_s-\bar{z}_i)^{-1}e^{-\varphi(z_i)-\tilde{\varphi}(\bar{z}_i)}e^{i(q+k)X(z_i,\bar{z}_i)}\nonumber\\[5 pt]
&&(\tilde{\varepsilon_s}\cdot k \tilde{\varepsilon_i}.\tilde{\psi}-\tilde{\varepsilon_i}\cdot\tilde{F_s}\cdot\tilde{\psi})(\bar{z}_i)(\varepsilon_s.k_i \varepsilon_i.\psi-\varepsilon_i.F_s.\psi)(z_i)+\cdots
\end{eqnarray}

here $F_s\equiv k_{s [\mu}\varepsilon_{s\nu]}$. we have written the contribution from $e^{i k X}$ and from $\partial X^{\mu}$ in a suggestive manner. The $\cdots$ indicates the terms which are sub-leading in the expansion of $|z_s-z_i|$. These subleading terms won't contribute when we integrate over the position of vertex operators.\\

 First, we will write the OPE in the real light cone coordinate.
 \begin{equation}
\begin{matrix}
    z = \sigma_{1} + i\sigma_{0} \\ 
    \bar{z} = \sigma_{1} - i\sigma_{0} 
\end{matrix}, \quad \Rightarrow \quad
\begin{matrix}
    z  \rightarrow z_{+ }= \sigma_{1} +\sigma_{0} \\ 
    \bar{z} \rightarrow z_{-} = \sigma_{1} - \sigma_{0} 
\end{matrix}
\end{equation}
 Then, we will follow the prescription as outlined in section \ref{hsz gauge}. This amounts to making the following substitution
\begin{itemize}
    \item $(z_s^+-z_i^+)^{k_s.k_i}(z_s^--z_i^-)^{ { -k_s.k_i}}$ will get modified to $e^{\frac{1}{\beta} z_{si+} \, \frac{k_s.k_i}{z_{si-}}}$
    \item $(z_s-z_i)^{-1}(\bar{z_s}-\bar{z_i})^{-1}$ will be modified to $\frac{1}{\beta z_{si-}^2}$
\end{itemize}
Here $z_{si+}\equiv z_s^+-z_i^+$ and $z_{si-}\equiv z_s^--z_i^-$.
After these modifications, the unintegrated OPE can be written as
\begin{eqnarray}
\label{momentum space ope}
V_s^{(0,0)}(z_s^+,z_s^-)V_h^{(-1,-1)}(z_i^+,z_i^-)\sim \mathrm{exp}(\frac{1}{\beta}\frac{k_s.k_i\,\, z_{si+}}{z_{si-}})\frac{1}{\beta z_{si-}^2}e^{-\varphi(z_i^+)-\tilde{\varphi}(z_i^-)}e^{i(k_s+k_i)X(z_i^+,z_i^-)}\nonumber\\
(\tilde{\varepsilon_s}.k_i \tilde{\varepsilon_i}.\tilde{\psi}-\tilde{\varepsilon_i}.\tilde{F_s}.\tilde{\psi})(z_i^-)(\varepsilon_s.k_i \varepsilon_i.\psi-\varepsilon_i.F_s.\psi)(z_i^+)+\cdots
\end{eqnarray}

The OPE for integrated vertex operators \footnote{Here both vertex operators are integrated. But after the soft limit, we will see that the hard operators don't need to be integrated.} can be written as
\begin{eqnarray}
\int dz^+_s dz^-_s V_s^{(0,0)}(z_s^+,z_s^-)\int d^2z_i V_h^{(-1,-1)}(z_i^+,z_i^-)\sim \int dz_s^+ dz_s^- d^2z_i \, \mathrm{exp}(\frac{1}{\beta}\frac{k_s.k_i\,\, z_{si+}}{z_{si-}})\frac{1}{\beta z_{si-}^2}e^{-\varphi(z_i^+)-\tilde{\varphi}(z_i^-)}\nonumber\\
e^{i(k_s+k_i)X(z_i^+,z_i^-)}(\tilde{\varepsilon_s}.k_i \tilde{\varepsilon_i}.\tilde{\psi}-\tilde{\varepsilon_i}.\tilde{F_s}.\tilde{\psi})(z_i^-)(\varepsilon_s.k_i \varepsilon_i.\psi-\varepsilon_i.F_s.\psi)(z_i^+)\nonumber\\
\end{eqnarray}
The $z_s^+$ integral can be done very easily. It will produce a delta function $\delta (\frac{k_s.k_i}{\beta z_{si-}})$. Then we can do the $z_s^-$ integral. 
\begin{eqnarray}
   && =\int dz_s^- d^2z_i\delta (\frac{k_s.k_i}{\beta z_{si-}})\frac{1}{\beta z_{si-}^2}e^{-\varphi(z_i^+)-\tilde{\varphi}(z_i^-)}e^{i(k_s+k_i)X(z_i^+,z_i^-)}
(\tilde{\varepsilon_s}.k_i \tilde{\varepsilon_i}.\tilde{\psi}-\tilde{\varepsilon_i}.\tilde{F_s}.\tilde{\psi})(z_i^-)(\varepsilon_s.k_i \varepsilon_i.\psi-\varepsilon_i.F_s.\psi)(z_i^+)\nonumber\\
&&=\int d^2z_i\frac{1}{k_s.k_i}e^{-\varphi(z_i^+)-\tilde{\varphi}(z_i^-)}e^{i(k_s+k_i)X(z_i^+,z_i^-)}
(\tilde{\varepsilon_s}.k_i \tilde{\varepsilon_i}.\tilde{\psi}-\tilde{\varepsilon_i}.\tilde{F_s}.\tilde{\psi})(z_i^-)(\varepsilon_s.k_i \varepsilon_i.\psi-\varepsilon_i.F_s.\psi)(z_i^+)\nonumber\\
\end{eqnarray}
 The numerator can be expanded out in the soft momenta $k_s$. The leading soft theorem (in the numerator we need to expand in $\mathcal{O}(k_s^0)$ order)
\begin{eqnarray}
 \int d^2 z_s  V_s^{(0,0)}(z_s^+,z_s^-)V_h^{(-1,-1)}(z_i^+,z_i^-)|_{\mathrm{leading}}&=&   \frac{1}{k_s.k_i}(\tilde{\varepsilon_s}.k_i \tilde{\varepsilon_i}.\tilde{\psi})(\varepsilon_s.k_i \varepsilon_i.\psi)e^{-\varphi(z_i^+)-\tilde{\varphi}(z_i^-)}e^{i(k_i)X(z_i^+,z_i^-)}\nonumber\\
 &=&\frac{(\tilde{\varepsilon_s}.k_i)(\varepsilon_s.k_i)}{k_s.k_i} V_h^{(-1,-1)}(z_i^+,z_i^-)\nonumber\\
\end{eqnarray}

The subleading soft theorem can be found by writing the numerator in $\mathcal{O}(k_s^1)$ 
\begin{eqnarray}
\big(i (k_s.X) (\tilde{\varepsilon_s}.k_i \tilde{\varepsilon_i}.\tilde{\psi})(\varepsilon_s.k_i \varepsilon_i.\psi)-(\varepsilon_s.k_i \varepsilon_i.\psi)(\varepsilon_i.\tilde{F_s}.\tilde{\psi})-(\tilde{\varepsilon_s}.k_i \tilde{\varepsilon_i}.\tilde{\psi})(\varepsilon_i.F_s.\psi)
\end{eqnarray}
The first term with $(k_s.X)(\varepsilon_s.k_i)$ can be written as a symmetric and anti-symmetric combination under the exchange of $k_s\leftrightarrow \tilde{\varepsilon_s}$. The symmetric part is BRST exact and won't contribute to the OPE. Only the anti-symmetric part will contribute. We can manipulate the above expression using the following identities
\begin{eqnarray}
\varepsilon_i.F_s.\psi=F_s^{\alpha\beta}\varepsilon_{i\alpha}\frac{\partial}{\partial \varepsilon_i^{\beta}} \varepsilon_i.\psi, \quad \tilde{F_s}^{\mu\nu} k_{i\nu} \frac{\partial}{\partial k_i^{\mu}} \tilde{\varepsilon_i}.\tilde{\psi} e^{i k_i.X}= i (X.k_s \tilde{\varepsilon_s}.k_i-X.\tilde{\varepsilon_s} k_s.k_i)\tilde{\varepsilon_i}.\tilde{\psi} e^{i k_i.X}
\end{eqnarray}
So, the subleading theorem can be written as
\begin{eqnarray}
 \int d^2 z_s   V_s^{(0,0)}(z_s)V_h^{(-1,-1)}(z_i)|_{\mathrm{subleading}}&=& \frac{1}{ k_s.k_i}\Bigg[(\varepsilon_s.k_i \tilde{\varepsilon_s}.k_i) k_{s}^{\alpha}\frac{\partial}{2 \partial k_i^{\alpha}}-(\varepsilon_s.k_i k_s.k_i) \tilde{\varepsilon_s}^{\alpha}\frac{\partial}{2 \partial k_i^{\alpha}}\nonumber\\
   &-&\varepsilon_s.k_i \tilde{F_s}^{\alpha\beta}(\tilde{\varepsilon_{i\alpha}}.\partial_{\tilde{\varepsilon_i}^{\beta}})+(\varepsilon \leftrightarrow \tilde{\varepsilon}) \Bigg] V_h^{(-1,-1)}(z_i^+,z_i^-)\nonumber\\
   &=& \frac{k_{i\alpha} e_{s}^{\alpha\rho}}{ k_s.k_i}\Bigg(k_s^{\beta} J_{\rho\beta}^{total}\,\, V_h^{(-1,-1)}(z_i^+,z_i^-)\Bigg)
\end{eqnarray}
Here $J+\tilde{J}\equiv J^{total}=k^{\mu} \frac{\partial }{\partial k_{\nu}}+\varepsilon^{\mu}\frac{\partial }{\partial \varepsilon_{\nu}}+\tilde{\varepsilon}^{\mu}\frac{\partial }{\partial \tilde{\varepsilon}_{\nu}}-(\nu \leftrightarrow \mu)$\\

The sub-subleading soft theorem can also be evaluated upon expanding the \footnote{The sub-subleading soft theorems are not universal. The non-universality comes from the non-minimal coupling- see \cite{Laddha:2017ygw}.}   the numerator in the $\mathcal{O}(k_s^2)$. The numerator can be written as
 \begin{eqnarray}
i (k_s.X) \Big((\tilde{\varepsilon_s}.k_i \tilde{\varepsilon_i}.\tilde{\psi})(\varepsilon_i.F_s.\psi)+(\varepsilon_s.k_i \varepsilon_i.\psi)(\tilde{\varepsilon_i}.\tilde{F_s}.\tilde{\psi})\Big)\nonumber\\
-(k_s.X)^2\Big(\frac{1}{2}(\tilde{\varepsilon_s}.k_i \tilde{\varepsilon_i}.\tilde{\psi})(\varepsilon_s.k_i\varepsilon_i.\psi)+\tilde{\varepsilon_i}.\tilde{F_s}.\tilde{\psi} \varepsilon_i.F_s.\psi\Big)
\end{eqnarray}
Doing the same manipulation as done above 
, one gets the sub-subleading soft theorem as 
\begin{eqnarray}
  \int d^2 z_s    V_G^{(0,0)}(z_s)V_G^{(-1,-1)}(z_i)|_{\mathrm{sub-subleading}}=\frac{e_s^{\alpha\beta}}{2 k_s.k_i}\Bigg(k_s. J_{\alpha}^{tot} k_s.J_{\beta}^{tot} \,\,V_G^{(-1,-1)}(z_i^+,z_i^-)\Bigg)
\end{eqnarray}
 This concludes our discussion of energetic soft theorems. In the next section, we will study the conformal soft theorem. 
 
\section{Conformal soft theorem}
\label{conformal soft theorem}
The scattering amplitudes in asymptotically flat space-time are usually written in momentum space. The incoming and outgoing states are plane waves (appropriate smearing) with fixed momenta. To understand the CFT dual of scattering in flat spacetime, it is convenient to write the amplitude in a celestial basis. We are interested in the scattering of massless particles, then the momenta and polarization of incoming/outgoing states can be written in celestial variables (For $D+2$ dimensional scattering, the celestial sphere is $D$ dimensional as (see appendix \eqref{identities})
\begin{eqnarray}
 &&   k^{\mu}(\omega_k,x)= \eta \omega_k (\frac{1+x^2}{2},x^a,\frac{1-x^2}{2}),  \quad \varepsilon_a^{\mu}(x)=\partial_a \hat{k}^{\mu}(x)=(x^a,\delta^{ab},-x^{a})
\end{eqnarray}
here $\eta=\pm 1$ represents the outcoming and incoming states. The polarization vector is $\varepsilon_{ab}^{\mu\nu}=\varepsilon_a^{\mu}\varepsilon_b^{\nu}$. On this basis, we diagonalize the boost and prepare a state as in boost-eigen states. This can be achieved by doing the Mellin transformation of incoming/outgoing states \cite{Pasterski:2017kqt,Pasterski:2017ylz}. In this celestial variable, the soft theorem becomes the ward identity of a current. The conformal operator/current has a discrete conformal dimension with $\Delta=1,0,-1,\cdots$. Hence, this is called the conformal soft theorem (see \cite{Pate:2019mfs,Donnay:2018neh,Adamo:2019ipt,Puhm:2019zbl,Guevara:2019ypd} for conformal soft theorems in gauge theory and gravity from amplitudes perspective). We are going to find these theorems directly from the world sheet of chiral strings.

\subsection{Conformal soft theorem from world-sheet}
The celestial vertex operators are defined as \footnote{here we are just doing the Mellin transformation with respect to $\omega$. This is enough for studying the conformal soft theorem, but one can also write the conformal operators with dependence only on the celestial variable. It requires writing the momenta and polarization on a celestial basis. } 
\begin{eqnarray}
 V_{\Delta_i}\equiv \int d^2 z_i  V_{\Delta_i}(z_i)=\int  d \omega_i \omega_i^{\Delta_i-1} \int d^2 z_i V(z_i)
\end{eqnarray}

We will start with the OPE two celestial vertex operators. The OPE in momentum space after all the world-sheet integral can be written as 
\begin{eqnarray}
\int d^2z_s  V_s^{(0,0)}(z_s^+,z_s^-)V_i^{(-1,-1)}(z_i^+,z_i^-)&\sim& \frac{1}{k_s.k_i}e^{-\varphi(z_i^+)-\tilde{\varphi}(z_i^-)}e^{i(k_s+k_i)X(z_i^+,z_i^-)}\nonumber\\
&&(\tilde{\varepsilon_s}.k_i \tilde{\varepsilon_i}.\tilde{\psi}-\tilde{\varepsilon_i}.\tilde{F_s}.\tilde{\psi})(\varepsilon_s.k_i \varepsilon_i.\psi-\varepsilon_i.F_s.\psi)\nonumber\\
&=& \frac{1}{\omega_s \omega_i\hat{k}_s.\hat{k}_i}e^{-\varphi(z_i^+)-\tilde{\varphi}(z_i^-)}e^{i(\omega_s \hat{k}_s+\omega_i \hat{k}_i)X(z_i^+,z_i^-)}\nonumber\\
&&(\omega_i\tilde{\epsilon_s}.\hat{k}_i \tilde{\epsilon_i}.\tilde{\psi}-\omega_s\tilde{\epsilon_i}.\tilde{F_s}.\tilde{\psi})(z_i^-)(\omega_i\epsilon_s.\hat{k}_i \epsilon_i.\psi-\omega_s\epsilon_i.F_s.\psi)(z_i^+)\nonumber\\
\end{eqnarray}
So, Now in the leading order in soft momenta $\omega_s$ we have 
\begin{eqnarray}
\int d^2z_s  V_s^{(0,0)}(z_s^+,z_s^-)V_i^{(-1,-1)}(z_i^+,z_i^-)\sim  \frac{\omega_i^2}{\omega_s \omega_i\hat{k}_s.\hat{k}_i}e^{-\varphi(z_i^+)-\tilde{\varphi}(z_i^-)}e^{i(\omega_i\hat{k}_i)X(z_i^+,z_i^-)}
(\tilde{\epsilon_s}.\hat{k}_i \tilde{\epsilon_i}.\tilde{\psi})(\epsilon_s.\hat{k}_i \epsilon_i.\psi)\nonumber\\
\end{eqnarray}
The OPE of conformal vertex operators can be evaluated as
\begin{eqnarray}
    V_{\Delta_s} V_{\Delta_i}\sim &&\int d\omega_s\,\, d\omega_i \,\, \omega_s^{\Delta_s-1}\,\,\omega_i^{\Delta_i-1}\int d^2z_s  V_G^{(0,0)}(z_s^+,z_s^-)V_G^{(-1,-1)}(z_i^+,z_i^-)\nonumber\\
&&\sim  \int d\omega_s\,\, d\omega_i \,\, \omega_s^{\Delta_s-2}\,\,\omega_i^{\Delta_i}e^{-\varphi(z_i^+)-\tilde{\varphi}(z_i^-)}e^{i(\omega_i\hat{k}_i)X(z_i^+,z_i^-)}
\frac{(\tilde{\epsilon_s}.\hat{k}_i\,\,\epsilon_s.\hat{k}_i)}{\hat{k}_s.\hat{k}_i}( \tilde{\epsilon_i}.\psi)( \epsilon_i.\psi)\nonumber\\
&&\sim  \int d\omega_s\,\, d\omega_i \,\, \omega_s^{\Delta_s-2}\,\,\omega_i^{\Delta_i}\frac{(\tilde{\epsilon_s}.\hat{k}_i\,\,\epsilon_s.\hat{k}_i)}{\hat{k}_s.\hat{k}_i}V_G^{(-1,-1)}(z_i^+,z_i^-)\nonumber\\
\end{eqnarray}
 Let's do the $\omega_s$ integral with $i \epsilon$ to make the integral convergent and after the integral, we will set it to zero.
 \begin{eqnarray}
     \int_0^{\infty} d \omega_s (\omega_s- i \epsilon)^{\Delta_s-2}= -\frac{(- i \epsilon)^{\Delta_s-1}}{ (\Delta_s-1)}
 \end{eqnarray}
 And $\lim_{\Delta_s \rightarrow 1}$ produces $\frac{-1}{\Delta_s-1}$ in the above equation.
And Mellin transforming the hard graviton vertex operator gives us $V_{\Delta_i+1}$. Hence the OPE becomes

\begin{eqnarray}
\lim_{\Delta_s\rightarrow 1}V_{\Delta_s}V_{\Delta_i}\sim\frac{1}{\Delta_s-1}\frac{(\tilde{\epsilon_s}.\hat{k}_i\,\,\epsilon_s.\hat{k}_i)}{\hat{k}_s.\hat{k}_i}V_{\Delta_i+1}
\end{eqnarray}

This is the leading soft theorem. The leading conformal soft operator can be defined as the residue of the operator at $\Delta_s=1$. For the sake of completion, we can write the conformal  graviton vertex operator as
\begin{eqnarray}\label{svt1}
  V_{1}^{\mathrm{soft}}=\mathrm{Res}_{\Delta \rightarrow 1}  V_{\Delta}&=&\mathrm{Res}_{\Delta \rightarrow 1}\int d^2z \, d \omega e_{\mu\nu}(i \partial X^{\mu}+ \omega k.\psi \psi^{\mu})(i \bar{\partial} X^{\nu}+\omega k.\tilde{\psi}\tilde{\psi}^{\nu}) \omega ^{\Delta-1}e^{i \omega  k.X}\nonumber\\
    &=&\mathrm{Res}_{\Delta \rightarrow 1}\int d^2z  \frac{\Gamma(\Delta)}{(-i \hat{k}.X)^{\Delta}}e_{\mu\nu} \Bigg[ (i \partial X^{\mu})( i \bar{\partial} X^{\nu}) +(\hat{k}.\psi \psi^{\mu})(i \bar{\partial} X^{\nu}) \frac{\Delta}{(-i \hat{k}.X)}\nonumber\\
&&+(i \partial X^{\mu})( \hat{k}.\tilde{\psi}\tilde{\psi}^{\nu})\frac{\Delta}{(-i \hat{k}.X)}
+(\hat{k}.\psi \psi^{\mu})(\hat{k}.\tilde{\psi}\tilde{\psi}^{\nu})\frac{(\Delta)(\Delta+1)}{(-i \hat{k}.X)^2}\Bigg]
\end{eqnarray}

 One needs to be very careful in taking the $\Delta \rightarrow 1$ residue. First one needs to do the OPE of soft vertex operators with other operators and then only take the residues at $\Delta=1$.  The leading soft theorem can be recast in terms of supertranslation mode at asymptotic infinity. The supertranslation mode is defined as \cite{Fotopoulos:2019vac}.
\begin{eqnarray}
    P(\mathrm{z_c,\bar{z}_c})\equiv \partial_{\mathrm{\bar{z}_c}}  V_{1}^{\mathrm{soft}}
\end{eqnarray}
here $\mathrm{z_c}$ is the coordinate on the celestial sphere. The celestial dependence of $V_{1}^{\mathrm{soft}}$ can be obtained by writing the momenta and polarization on a celestial basis. \\

On the celestial sphere $V_{1}^\mathrm{{soft}}\mathrm{(z_c,\bar{z}_c)}$ has $(\Delta,J)=(1,2)$. 
The supertranslation operator $\mathrm{P}(\mathrm{z_c,\bar{z}_c})$ has $(\Delta,J)=(2,1)$ and $SL(2,C)$ weight as $(h,\bar{h})=(\frac{3}{2},\frac{1}{2})$.\\

Now we will move on to the sub-leading theorem. Following the same logic as above, We have  
\begin{eqnarray}
 V_{\Delta_s} V_{\Delta_i}&&\sim\int d\omega_s\,\, d\omega_i \,\, \omega_s^{\Delta_s-1}\,\,\omega_i^{\Delta_i-1}\int d^2 z_s  V_s^{(0,0)}(z_s^+,z_s^-)V_i^{(-1,-1)}(z_i^+,z_i^-)\nonumber\\
&&\sim \int d\omega_s\,\, d\omega_i \,\, \omega_s^{\Delta_s-1}\,\,\omega_i^{\Delta_i-1}\frac{ \hat{k}_{i\mu}e_s^{\mu\rho}}{ \hat{ k}_s.\hat{k}_i}\Big( \hat{k}_s^{\nu}J_{\rho\nu}^{total}\,\,V_i^{(-1,-1)}(z_i^+,z_i^-)\Big)\nonumber\\
&&= \frac{1}{\Delta_s} \frac{ \hat{k}_{i\mu}e_s^{\mu\rho}}{  \hat{k}_s.\hat{k}_i}\Big( \hat{k}_s^{\nu}J_{\rho\nu}^{total}\,\, V_i^{(-1,-1)}\Big)\nonumber\\
\end{eqnarray}

The sub-leading soft theorem can be understood as superrotation mode at asymptotic infinity. The corresponding conformal soft operator can be defined as the Residue at $\Delta \rightarrow 0$.
\begin{eqnarray}\label{svt2}
 V_{0}^{soft}\equiv    \mathrm{Res}_{\Delta\rightarrow 0}V_{\Delta}    &=&\mathrm{Res}_{\Delta\rightarrow 0}\int d^2z  \frac{\Gamma(\Delta)}{(-i k.X)^{\Delta}}e_{\mu\nu} \Bigg[ (i \partial X^{\mu})( i \bar{\partial} X^{\nu}) +(k.\psi \psi^{\mu})(i \bar{\partial} X^{\nu}) \frac{\Delta}{(-i k.X)}\nonumber\\
&&+(i \partial X^{\mu})( k.\tilde{\psi}\tilde{\psi}^{\nu})\frac{\Delta}{(-i k.X)}
+(k.\psi \psi^{\mu})(k.\tilde{\psi}\tilde{\psi}^{\nu})\frac{(\Delta)(\Delta+1)}{(-i k.X)^2}\Bigg]
    \end{eqnarray}
Again, we need to be careful about the residue being taken. First one should do the OPE of the soft operator with other hard operators then only the residue should be taken. 

We will end this section with the sub-subleading soft theorem. This theorem can be expressed as a pole in the $\Delta\rightarrow -1$ limit. 


\begin{eqnarray}
V_{\Delta_s}V_{\Delta_i} &&\sim \int d\omega_s\,\, d\omega_i  \omega_s^{\Delta_s-1}\,\,\omega_i^{\Delta_i-1}\int d^2 z_s V_s^{(0,0)}(z_s^+,z_s^-)V_i^{(-1,-1)}(z_i^+,z_i^-)\nonumber\\
&&\sim \int d\omega_s\,\, d\omega_i  \omega_s^{\Delta_s-1}\,\,\omega_i^{\Delta_i-1}\frac{ e_s^{\mu\nu }\omega_s}{ \omega_i\hat{ k}_s.\hat{k}_i}\Big(\hat{k}_s.J_{\mu}^{total}\hat{k}_s.J_{\nu}^{total}\,\, V_{i}^{(-1,-1)}(z_i^+,z_i^-)\Big)\nonumber\\
&&\sim \frac{1}{\Delta_s+1}\frac{ e_s^{\mu\nu }}{ \hat{k}_s.\hat{k}_i}\Big( \hat{k}_s.J_{\mu}^{total}\hat{k}_s.J_{\nu}^{total}\,\,V_{\Delta_i}^{(-1,-1)}\Big)\nonumber
\end{eqnarray}

In a similar manner, the corresponding conformal operator is defined as the Residue at $\Delta \rightarrow -1$. These soft theorems get generalized to the whole tower with $\Delta=1,0,-1,-2,\cdots$. This tower generates the $w_{1+\infty}$ algebra. These currents are generated by the asymptotic symmetry at the conformal boundary.

\section{Stress tensor}
\label{stress tensor}
The statement of soft theorems becomes the ward identity of the celestial sphere. In particular, the sub-leading soft theorem gives an identity similar to the ward identity of stress tensor after doing the shadow transform \cite{Kapec:2016jld}. We can get the celestial stress tensor from the world sheet as well \footnote{On the world sheet we have a stress tensor. The celestial stress tensor is different from the world-sheet ones. The relationship between them is not known to us. }. The chiral string has the same critical dimension as usual string theory namely $D=10, 26$ for superstring and bosonic string respectively. But tree amplitudes can be consistently written in any dimension. We will be interested in $d=4$ where we will make contact with the celestial sphere parametrized by $\mathrm{z},\bar{\mathrm{z}}$.\\
Let's parametrize the polarization and momenta of the vertex operators as
\begin{eqnarray}
  &&  p^{\mu}=\frac{1}{\sqrt{2}}\omega\Big(1+\mathrm{z}\bar{\mathrm{z}}, \mathrm{z}+\bar{\mathrm{z}},-i (\mathrm{z}-\bar{\mathrm{z}}),1- \mathrm{z} \bar{\mathrm{z}}\Big)\nonumber\\
   && \varepsilon_{\mu}^{(+)}(p)=\frac{1}{\sqrt{2}}(-\bar{\mathrm{z}},1,-i,-\bar{\mathrm{z}}),\quad  \varepsilon_{\mu}^{(-)}(p)=\frac{1}{\sqrt{2}}(-\mathrm{z},1,i,-\mathrm{z})\nonumber\\
\end{eqnarray}
The soft operators for leading, sub-leading, and sub-subleading theorems in celestial basis are written in \cite{Freidel:2021dfs,Kapec:2016jld}. The leading and subleading soft operators are
\begin{eqnarray}
 && \mathrm{  S_+^{(0)}=-\frac{1}{2 \omega} \sum_{k=1}^n \omega_k \frac{\bar{\mathrm{z}}-\bar{\mathrm{z}}_k}{\mathrm{z}-\mathrm{z}_k}\qquad \qquad \qquad S_-^{(0)}=-\frac{1}{2 \omega} \sum_{k=1}^n \omega_k \frac{\mathrm{z}-\mathrm{z}_k}{\bar{\mathrm{z}}-\bar{\mathrm{z}}_k}}\nonumber\\
  && \mathrm{ S_+^{(1)}=\frac{1}{2} \sum_{k=1}^n \frac{(\bar{\mathrm{z}}-\bar{\mathrm{z}}_k)^2}{\mathrm{z}-\mathrm{z}_k}\Big[\frac{2 \bar{h}_k}{\bar{\mathrm{z}}-\bar{\mathrm{z}}_k}-\partial_{\bar{\mathrm{z}}_k}\Big]\quad  S_-^{(1)}=\frac{1}{2} \sum_{k=1}^n \frac{(\mathrm{z}-\mathrm{z}_k)^2}{\bar{\mathrm{z}}-\bar{\mathrm{z}}_k}\Big[\frac{2 h_k}{\mathrm{z}-\mathrm{z}_k}-\partial_{\mathrm{z}_k}\Big]}
\end{eqnarray}
In both of these soft operators, $\mathrm{k}$ labels the hard particles with the celestial label as $\mathrm{z_k}$, and soft particles have celestial coordinates as $\mathrm{z}$. 

For the stress tensor, One needs the sub-leading soft operator. This is established by asymptotic symmetry analysis for gravitons \cite{Kapec:2016jld}. The corresponding ward identity for this symmetry (negative helicity gravitons) leads to 
\begin{eqnarray}
\mathrm{\langle out|N_{\bar{\mathrm{z}}\bar{\mathrm{z}}}^{(1)}\mathcal{S}|in\rangle=\frac{1}{2} \sum_{k=1}^n \frac{(\mathrm{z}-\mathrm{z}_k)^2}{\bar{\mathrm{z}}-\bar{\mathrm{z}}_k}\Big[\frac{2 h_k}{\mathrm{z}-\mathrm{z}_k}-\partial_{\mathrm{z}_k}\Big]\langle out|\mathcal{S}|in \rangle}
\end{eqnarray}
On the LHS, we have amplitudes of outgoing, ingoing, and one extra subleading-soft graviton (negative helicity). On the RHS, we have soft factors and amplitudes of outgoing and ingoing states. Here  $\mathcal{S}$ is the S- matrix between incoming and outgoing states and $\mathrm{N}_{\bar{\mathrm{z}}\bar{\mathrm{z}}}^{(1)}$ is the extra graviton. If we do the shadow transform of the soft graviton operator then, the RHS of the above equation looks like the ward identity for the stress tensor. 
\begin{eqnarray}
\label{stress-tensor}
\mathrm{T_{\mathrm{zz}}(\mathrm{z,\bar{z}})\equiv \frac{3}{\pi }\int d^2\mathrm{w} \frac{1}{\mathrm{(z-w)^4}}\mathrm{N_{\bar{w}\bar{w}}^{(1)}(w,\bar{w})}}
\end{eqnarray}
After the shadow transform, the insertion of the stress tensor in the correlation function/amplitude can be written as
\begin{eqnarray}
\label{stress tensor insertion}
   \mathrm{ \langle T(z,\bar{z}) \prod_{i=1}^n \phi_i(z_i,\bar{z}_i) \rangle= \sum_k \Bigg(\frac{h_k}{(z-z_k)^2}+\frac{1}{z-z_k} \frac{\partial}{\partial z_k}\Bigg)\langle \prod_{i=1}^n \phi_i(z_i,\bar{z}_i) \rangle}
\end{eqnarray}
here $\mathrm{\phi_i}$ can be any other incoming/outgoing states in a celestial basis.\\

With this preparation for chiral string, the celestial stress tensor is just the shadow transform of the sub-leading soft graviton vertex operator.\footnote{ For the negative helicity vertex operator, we have $\Delta=0$ and spin $\mathrm{h-\bar{h}=-2}$. After the shadow transform the resulting operator will have dimension $(2,0)$. A similar result can be obtained for positive helicity graviton.} The sub-leading soft operator (here we need celestial weight as $\mathrm{(h,\bar{h})=(-1,1)}$ is
\begin{eqnarray}
  \mathrm{  V_{G}^{(h=-1,\bar{h}=1)}}(\mathrm{z,\bar{z}})&=&\mathrm{Res}_{\Delta\rightarrow 0}\int dz d\bar{z} \frac{\Gamma(\Delta)}{(-i k_s.X)^{\Delta}}e_{\mu\nu}(\mathrm{z,\bar{z}}) \Bigg[ (i \partial X^{\mu})( i \bar{\partial} X^{\nu}) +(k_s.\psi \psi^{\mu})(i \bar{\partial} X^{\nu}) \frac{\Delta}{(-i k_s.X)}\nonumber\\
&&+(i \partial X^{\mu})( k_s.\tilde{\psi}\tilde{\psi}^{\nu})\frac{\Delta}{(-i k_s.X)}
+(k_s.\psi \psi^{\mu})(k_s.\tilde{\psi}\tilde{\psi}^{\nu})\frac{(\Delta)(\Delta+1)}{(-i k_s.X)^2}\Bigg]
    \end{eqnarray}
We can write the momenta $k_s$ and appropriate polarization $e_{\mu\nu}$ in terms of a celestial variable. But this is not necessary here. After doing the required shadow transform we have the celestial stress tensor as
\begin{eqnarray}\label{celest}
    \mathrm{T_{zz}(z,\bar{z})= \frac{3}{\pi} \int dw d \bar{w}\frac{1}{(z-w)^4} V_{G}^{h=-1,\bar{h}=1}(w,\bar{w})}
\end{eqnarray}
This stress tensor when inserted into the correlation function produces exactly the \eqref{stress tensor insertion}. This should have already been anticipated by our discussion on soft theorems.  The anti-holomorphic stress tensor can also be defined using the positive helicity graviton and then doing the shadow transform. The main point that we want to emphasize in this section is that we have a well-defined notion of stress tensor written in terms of field variables of the worldsheet theory. We can use this stress tensor of \eqref{celest} and compute the OPE
 with other vertex operators that have celestial conformal dimension $\Delta$ without relying on the collinear limit as often done in celestial holography. The calculation of the OPE goes similarly as done in section \ref{energetic soft theorem}. A very concrete and different calculation that one can do is to find the OPE of two stress tensors ($\mathrm{TT}$). This is left for future works (see \cite{Banerjee:2022wht} for subtlety in taking the consecutive soft limits). One can define a similar stress tensor in any world-sheet theory like string theory, ambi-twistor string, etc. that gives scattering amplitude in target space. \\

\section{String world-sheet and celestial CFT}
\label{worldsheet and CCFT}

In section \ref{energetic soft theorem}, we demonstrated that in chiral string theory, the leading and subleading soft terms come from the contribution when soft and hard vertex operators collide. This means that the soft theorem is determined by the form of the OPE between soft and hard vertex operators. We expect that this result is more general. In this section, we would also investigate OPEs of CCFT themselves. The CCFT OPEs come from the collinear limit of the amplitudes. Since we know that at the leading order, the collinear limit and soft limit coincide, we should also expect that we could derive it directly from the OPEs of vertex operators. 

In this section, we would like to put the relationship between chiral strings and CCFT into a more general and abstract setting. Instead consider particular models or chiral strings, we consider a generic world-sheet CFT with a field theory description.
When we say ``a generic world-sheet CFT with a field theory description'',  we mean two things.
\begin{enumerate}

    \item There is a 2D CFT with its primary field content $\{\mathcal{O}^{i}\} = \{V^{i}_k, \, \mathcal{O}^{(a)}\}$, where $V_{k}^{i}$ are the ``vertex operators'', they are parameterized by $D$ continuous variables collectively labeled as $k$, which is interpreted as the `` momentum '' of the particle, while the discrete index $i$ labels all other quantum numbers collectively, such as spin and particles types, etc. The on-shell condition demands that they have dimensions $(1, 1)$. 
    \item There is a one-to-one correspondence between the one-particle spectrum of the field theory $|k; i\rangle $ and the (bare) vertex operators $V^{i}_{k}$. The (formal) prescriptions of tree amplitudes are as expected:
    \begin{equation}
        A^{i_{1}\cdots i_{n}}{}(k_{1}, \cdots k_{n}) = \frac{1}{\mathcal{V}_{\textrm{CKG}}}\int dm \left\langle \, V_{k_{1}}^{i_{1}}\cdots V_{k_{n}}^{i_{n}}\, \right\rangle_{S^{2}},
    \end{equation}
    where the integral integrates over the moduli space of $n-$punctured sphere. Since CFT on a sphere still possesses residual symmetries, i.e. the conformal Killing group,  one needs to divide the integral by the volume of such group ($\mathcal{V}_{\textrm{CKG}}$).
\end{enumerate}

In such a world-sheet CFT, one needs to gauge fix the conformal Killing symmetries by fixing the positions of three vertex operators. In order the evaluate the amplitude explicitly, one needs to apply BRST formalism, and extend the Hilbert space of the original world-sheet CFT by introducing ghosts and anti-ghosts. We could also construct the BRST charges $Q_{\textrm{BRST}}$, and for each vertex operator $V_{k}^{i}$, we could construct integrated vertex operators $U_{k}^{i}$ and unintegrated vertex operator $W_{k}^{i}$. Using this we could calculate the gauged fixed tree amplitude. 
\begin{equation}\label{eq:6-2}
         A^{i_{1}\cdots i_{n}}{}(k_{1}, \cdots k_{n}) = \int_{\mathbb{C}}\prod_{s=4}^{n}d^{2}z_{i}\left\langle U^{(i_{s}}(z_{s}) \, W^{i_{1}}_{k_{1}}(\hat{z}_{1}) W^{i_{2}}_{k_{2}}(\hat{z}_{2})W^{i_{3}}_{k_{3}}(\hat{z}_{3})\right\rangle_{S^{2}},
\end{equation}
    where we gauge fixed the positions of three vertex operators to  $\hat{z}_{1}, \hat{z}_{2}, \hat{z}_{3}$.

Using the BRST charge, one can single out the physical states of the full Hilbert space as the cohomology of $Q_{\textrm{BRST}}$\footnote{Recall that the cohomology of a Hilbert space are states that are closed ($Q_{\textrm{BRST}}\mathcal{O} = 0$), but are not exact ($\mathcal{O} = Q_{\textrm{BRST}}\mathcal{O}'$).}, where each $V_{k}^{i}$ correspond to a representative of the BRST cohomology class. One can choose a suitable basis for the full Hilbert space such that all the non-closed states are orthogonal to the closed state. (i.e. The inner products of any physical state with any non-physical states are 0.)

Consider the OPE coefficients of the two integrated vertex operators, it can be written as:\footnote{The OPEs of two integrated vertex operators cannot contain unintegrated vertex operators $W_{k}$, since the violates ghost charge conservation; $U_{k}$ have ghost charge 0 while $W_{k}$ have ghost charge 1. }
\begin{equation}\label{eq:3.3}
    U^{i}_{p}(0)U^{j}_{q}(z) = \sum_{k} C^{ij}{}_{k}(z, \partial, \bar{\partial})U^{k}_{p+q}(0) + \sum_{\alpha} \gamma_{ij\alpha}(z, \partial, \bar{\partial})\mathcal{O}^{\alpha}(0).
\end{equation}
The OPEs are organized in a way that the descendants are represented as the $\partial, \bar{\partial}$ dependence of the coefficients $C^{ijk}$. We separated the BRST closed vertex operators $U^{i}$ and the non-closed primaries $\mathcal{O}^{(a)}$. The contribution of the non-closed operators could be ignored when evaluating this equation inside correlation functions, as the closed and non-closed operators are orthogonal.

From the conformal invariance of general 2D CFT, the OPE coefficients can be written as
\begin{equation}
C^{ij}{}_{k}(z, \partial, \bar{\partial}) = c^{ij}{}_{k}z^{h_{ijk}}\bar{z}^{\bar{h}_{ijk}}\left( 1 -\frac{h_{ijk}}{h_{k}} z\partial + \cdots \right)\left( 1 - \frac{\bar{h}_{ijk}}{h_{k}} \bar{z}\bar{\partial} + \cdots \right),
\end{equation}
where $h_{ijk} = h_{k} - h_{i} - h_{j}$ and $h_{i}, h_{j}, h_{k}$ is the holomorphic dimension for  $U^{i}, U^{j}$ and $U^{k}$ respectively. The holomorphic dimensions of $U^{k}$ could be written as  $h = N \pm k^{2}, \bar{h} = M \pm k^{2}$, where $N, M$ are some integers. For the integrated vertex operators to make sense, we need to demand that its holomorphic dimensions equal $(1, 1)$, and this will give us the on-shell conditions for the momentum. $k^{2} = - M^{2}$, where $M^{2} = N-1$ is the masses of the corresponding particle.
Since $U_{p}^{i}$ and $U_{q}^{j}$ are on shell, so they have dimensions $(1, 1)$. While $h_{k} = 1 \pm \left(\,(p+q)^2 + M_{k}^{2}\right)$, $M_{k}$ is the mass correspond to $U^{k}$. Thus
\begin{equation}\label{onshellcond}
  h_{ijk} = \pm\left(\,(p+q)^{2} + M_{k}^{2}\,\right) -1 , \quad     \bar{h}_{ijk} = \pm\left(\,(p+q)^{2} + M_{k}^{2}\,\right) -1.
\end{equation}
When the sign in the expression of  $h_{ijk}$ and $\bar{h}_{ijk}$ are the same we have an ``usual string theories'', while when the sign is different, we have a ``chiral strings-like theoreis''.

Our strategy for deriving the OPEs of the CCFT is to first calculate the world-sheet OPEs of two vertex operators inside the correlators, and evaluate the integration of the moduli space   (\ref{eq:6-2}). Note that the contribution from the descendants in the OPE vanishes upon integration as it is a total derivative. Assuming the that the descendants equals to $\partial \mathcal{O}$, then
\begin{align*}
    &\int d^{2}z_{1}d^{2}z_{2}(z_{1} - z_{2})^{\alpha}\partial_{z_{2}}\mathcal{O}(z_{2})  = \int d^{2}z_{1}d^{2}z_{2}\partial_{z_{2}}(z_{1} - z_{2})^{\alpha}\mathcal{O}(z_{2}) \\
    =& -(-1)^{\alpha} \int d^{2}z_{1}d^{2}z_{2}\partial_{z_{1}}(z_{1} - z_{2})^{\alpha}\mathcal{O}(z_{2}) = 0.\\
\end{align*}
Thus, we have
\begin{align}\label{generalope}
    &\int d^{2}z_{1}d^{2}z_{2}\cdots \; \langle U^{i}_{p}(z_{1})U^{j}_{q}(z_{2}) \cdots \rangle  =\sum_{k}\int d^{2}z_{1}d^{2}z_{2}\cdots \; \bar{z}_{12}{}^{\bar{h}_{ijk}} z_{12}{}^{h_{ijk}}  \,  c^{ij}{}_{k}\langle U^{k}_{p + q}(z_{2}) \; \cdots \rangle.
\end{align}
Therefore, the only information we need to get the explicit form of the OPE is the coefficient $c^{ij}{}_{k}$.

We could get $c^{ij}{}_{k}$ by the 3-point correlators of the vertex operators. In our case, it is just the 3-point amplitude:
\begin{equation}
    \langle U^{\alpha}(z_{1})U^{\beta}(z_{2})U^{\gamma}(z_{3}) \rangle = \frac{c^{\alpha\beta\gamma}}{|z_{12}|^{2}|z_{13}|^{2}|z_{23}|^{2}},
\end{equation}
where we have used the short-hand notation where the Greek indices represent both momentum and spin indices: $\alpha = (i, p), \beta = (j, q), \gamma = (k, r)$;
\begin{equation}
    c^{\alpha \beta \gamma} = A^{ijk}(p, q , r)\delta^d\left(p+q+r\right), \quad \Rightarrow c^{ijk} = A_{3}^{ijk}(p, q , -p-q).
\end{equation}

Note that we also have two point correlators of the vertex operators:
\begin{equation}
\langle U^{\alpha}(z_{1}) U^{\beta}(z_{2})\rangle  = \frac{\kappa^{\alpha\beta}}{|z_{12}|^{2}}, \quad \kappa^{\alpha\beta} = \kappa^{ij}\delta^d(p+q).
\end{equation}
Conventionally, we can always choose a basis for the primary operators such that their two-point correlators remain diagonal. But in order not to obscure the fact that these vertex operators correspond to the physical spectrum of the target space, we will not do that here.

$\kappa^{ij}$ can be constructed from the polarization tensors and it depends on what kind of polarization basis one chooses. $\kappa^{\alpha\beta}$ can be used as a metric for the space of primaries, and it can be used to lower and raise the indices of the vertex operators.
\begin{equation}
\kappa_{\alpha\beta} \equiv (\kappa^{\alpha\beta})^{-1}.\quad U_{\alpha} = \kappa_{\alpha\beta}U^{\beta} ,\quad \Rightarrow U_{i}(p; z) = \kappa_{ij}U^{j}(-p; z).
\end{equation}

In the OPE equation (\ref{eq:3.3}), the primaries are raised on the right-hand side. If one is not careful about the fact that the vertex operators are not diagonal with respect to the two-point correlators, they may think the vertex operators that appeared on the right-hand side should have momentum $-p-q$ instead of $p+q$ according to the momentum conservation, which contradicts with direct calculation. 

In the limit when $(p + q)$ approaches on shell, from (\ref{onshellcond}), $h_{ijk}$ and $\bar{h}_{ijk}$ approaches $-1$. In particular, if we use the limit $\lim_{\epsilon \rightarrow 0_{+}} |x|^{-1 + \epsilon} = \frac{1}{\epsilon}\delta(x)$, the double integral in (\ref{generalope}) will reduce to a single integral. We get

\begin{align}
    &\int d^{2}z_{1}d^{2}z_{2}\cdots \; \langle U^{i}_{p}(z_{1})U^{j}_{q}(z_{2}) \cdots \rangle  \sim \frac{1}{p\cdot q + M_{k}^{2}} \int d^{2}z_{1}\cdots  \;  c^{ij}{}_{k}\langle U^{k}_{p + q}(z_{2}) \; \cdots \rangle.
\end{align}
Since we are interested in only the massless particles (gravitons), the equation above is essentially the leading collinear limit or the leading soft limit. Since one could understand the $p\cdot q  = \frac{1}{2}\omega_{p}\omega_{q}(x_{p} - x_{q})^{2} \rightarrow 0$ as $\omega_{q}\rightarrow 0$ or $x_{p}\rightarrow x_{q}$.

Therefore, we can write the relations between graviton vertex operators.
\begin{equation}\label{zope}
    Z^{i}(p)Z^{j}(q) \sim \frac{A^{ijk}(p, q , -p-q)}{2p\cdot q}Z_{k}(p + q),\quad Z^{i}(p)\equiv \int d^{2}z U^{i}_{p}(z) , \quad p\cdot q\rightarrow 0.
\end{equation}
Of course, these equalities only make sense when they are sandwiched between correlators. 
We have identified the collinear splitting function.
\begin{equation}
    \mathrm{Split}(p, q)=\frac{A^{ijk}(p, q, -p -q)}{2 p\cdot q}
\end{equation}
Using the splitting function, one can easily derive the leading order OPEs in the CCFT, and derive the infinite tower of conformal soft theorems as usual.

For example, in the closed  (chiral) string theory, the three-point amplitudes of graviton can be written as (see \cite{Polchinski:1998rr,Polchinski:1998rq} for explicit factors and normalization )
\begin{equation}
A_{GGG}= \delta^d(k_1+k_2+k_3)e_{1\mu\nu}e_{2\alpha\beta}e_{3\rho\sigma} V^{\mu\alpha\rho} V^{\nu\beta\sigma}
\end{equation}
and
\begin{equation}
V_{\mu\alpha\rho}=k^{\mu}_{23}\eta^{\alpha\rho}+k^{\alpha}_{31}\eta^{\mu\rho}+k^{\rho}_{12}\eta^{\alpha\mu}, \quad k_{ij}=k_i-k_j.
\end{equation}

We can chose the polarization as $e_i^{\mu\nu}=\varepsilon_{a_i}^{\mu}\varepsilon^{\nu}_{\tilde{a}_i}$. Then the amplitudes can be written in the factorized form as
\begin{equation}
    A_{GGG}=\hat{A}_3(\omega_i,\varepsilon_{ai}) \hat{ A}_3 (\omega_i,\varepsilon_{\tilde{a}i})
\end{equation}
 
The $\hat{A}_3(\omega_i,\varepsilon_{ai})$ can also be written in celestial variable (see appendix \ref{identities} and \cite{Jiang:2021csc} for more details ). 
\begin{equation}
    \hat{A}_3(\omega_i,\varepsilon_{ai})= 2 \Bigg[\omega_2 x_{21}^{a_1} \delta^{a_2a_3}-\omega_1 x_{12}^{a_2} \delta^{a_1 a_3}+\frac{\omega_1\omega_2}{\omega_1+\omega_2} x_{12}^{a_3} \delta^{a_1 a_2}\Bigg].
\end{equation}
The splitting function can be found very easily by
 \begin{equation}
     \mathrm{Split}(k_1, k_2)=\frac{A_{GGG}}{2 k_1\cdot k_2}=-\frac{\hat{A}_3(\omega_i,\varepsilon_{ai}) \hat{ A}_3 (\omega_i,\varepsilon_{\tilde{a}i})}{(x_{12})^2 \omega_1 \omega_2} 
 \end{equation}
 One can Mellin transform these functions to write in a boost eigen basis. We can compare $\hat{A}_3(\omega_i,\varepsilon_{ai})$ with expression in \eqref{w ope} in the parenthesis. Hence, after doing the Mellin transformation it will reproduce the result in \eqref{w ope final}. If one uses the $4$-d momenta and polarization basis ($\pm$ helicity basis) then it will reproduce the result of $\eqref{w ope final 4d}$.

    

\section{Summary and Discussions}
\label{Discussions}
In this article, we have studied chiral string theory and its relationship with celestial holography. In section \ref{energetic soft theorem}, we studied the soft theorem in chiral string formalism. We saw that the leading and subleading soft theorem can be derived completely from the OPEs of vertex operators.  For the bosonic case, we proved the universality of soft theorems. For the supersymmetric case, we only considered the soft theorems of gravitons in the HSZ gauge.
 
In section \ref{conformal soft theorem},  we studied the conformal soft theorems terms of conformal vertex operators, where we change the momentum of vertex operators into conformal basis. Using these conformal vertex operators, one could write down the conformal soft limit and define the conformal soft vertex operators (\ref{svt1}), (\ref{svt2}).  It could be generalized to the whole tower of conformal soft theorems realizing $w_{1+\infty}$ symmetry algebra. In section \ref{stress tensor}, we have defined a celestial stress tensor in terms of the soft vertex operators in (\ref{celest}) In the last section \ref{worldsheet and CCFT}, We study the relationship between CCFT and usual world-sheet CFT in more general settings. We find that under fairly general circumstances, one can derive leading order OPEs for CCFT directly from the world-sheet vertex operators. 

There are several future directions we would wish to pursue.
\begin{itemize}
    \item In the future, we wish to study the implications of unitarity, causality as well as analyticity of the CCFT correlation function from the chiral string. We believe it will serve as a good stepping stone to understanding the celestial CFT.
    \item  The immediate question one can ask is what happens at the loop level. How one can organize the celestial CFT at the loop level. From the chiral string perspective, one needs to find the propagator on the torus and study the torus amplitude. One can try to find the central charge of the celestial CFT using the OPE of the stress tensor that we have provided, and see if loop corrections will have a correction on it.
    \item Another interesting direction is to understand more about the CCFT OPEs using the vertex operators. As equations among the vertex operators like (\ref{drope}) don't refer to amplitudes with specific numbers of particles. Making various statements much more general than statements on the amplitude level.
    \item We would also like to understand the interplay between asymptotic symmetries and chiral string theories. As the soft theorems are the Ward identities for those symmetries, which could be derived from the chiral strings OPEs, maybe there is a more direct way to see those symmetries in terms of vertex operators. 
\end{itemize} 

\begin{acknowledgments}

We are indebted to Y.~Li, M.~Ro\v cek for helpful discussions, and Y.Li for his initial collaboration on the project.

\end{acknowledgments}
\appendix

\section{Kinematics and celestial sphere}
\label{identities}
In this section, we will consider the kinematics of  $D+2$ dimensional spacetime which can be described by coordinates on the $D$ dimensional celestial sphere. First, we will start with null momentum (which is relevant for massless particles). It can be parametrized by
\begin{eqnarray}
 &&   k^{\mu}(\omega_k,x)= \eta \omega_k (\frac{1+x^2}{2},x_1^a,\frac{1-x^2}{2}), \qquad n^{\mu}= (-1,0,1),
\end{eqnarray}
here $\eta=\pm 1$ represents the outcoming and incoming states. Now, we will introduce the polarization vector and its contractions ($\mu=0,\cdots D+1$ and $a=1,\cdots D$) as
\begin{eqnarray}
 &&  \varepsilon_a^{\mu}(x)=\partial_a \hat{k}^{\mu}(x)=(x^a,\delta^{ab},-x^{a})\nonumber\\
&&n.n=\hat{k}.\hat{k}=\varepsilon_a.n=\varepsilon_a.\hat{k}=0,\quad n.\hat{k}=1, \varepsilon_a.\varepsilon_b=\delta_{ab}, \quad \varepsilon_{ab}^{\mu\nu}=\varepsilon_a^{\mu}\varepsilon_b^{\nu}
\end{eqnarray}
The product of momenta and polarization can also be written in celestial variables.
\begin{eqnarray}
    \hat{k}_i.\hat{k}_j=-\frac{1}{2}x_{ij}^2, \quad \hat{k}_i.\varepsilon_a(x_j)= x_{ij}^a, \quad \varepsilon_a(x_i).\varepsilon_b(x_j)=\delta^{ab}
\end{eqnarray}
The product of two polarizations $\varepsilon_a^{\mu}\varepsilon_b^{\nu}$(which is going to be useful for gravitons) can be decomposed into the representation of $SO(D)$ as 
\begin{eqnarray}
    D\otimes D= \frac{(D+2)(D-1)}{2}+\frac{D(D-1)}{2}+1
\end{eqnarray}
They will correspond to the graviton/antisymmetric B field and dilaton respectively. 

For 4d spacetime, the Celestial sphere is 2-dimensional. We can use complex coordinates to specify it.
\begin{eqnarray}
    x^1=\frac{\mathrm{z}+\bar{\mathrm{z}}}{2}, \qquad x^2=\frac{-i(\mathrm{z}+\bar{\mathrm{z}})}{2}, \quad \hat{k}^{\mu}=\frac{1}{2}(1+ \mathrm{z}\bar{\mathrm{z}},\mathrm{z}+\bar{\mathrm{z}},-i (\mathrm{z}-\bar{\mathrm{z}}),1-\mathrm{z}\bar{\mathrm{z}})
\end{eqnarray}
The polarization vectors in the helicity basis are
\begin{eqnarray}
    \varepsilon_+^{\mu}=\frac{1}{\sqrt{2}} (\varepsilon_1^{\mu}- i \varepsilon_2^{\mu})=\frac{1}{\sqrt{2}} (\mathrm{\bar{z}},1,-i,-\bar{\mathrm{z}}), \quad  \varepsilon_-^{\mu}=\frac{1}{\sqrt{2}} (\varepsilon_1^{\mu}+ i \varepsilon_2^{\mu})=\frac{1}{\sqrt{2}} (\mathrm{z},1,-i,-\mathrm{z}), 
\end{eqnarray}
These polarization satisfy $ \varepsilon_+.\varepsilon_+=\varepsilon_-.\varepsilon_-=0, \varepsilon_-.\varepsilon_+=1$. For gravitons we need
\begin{eqnarray}
\varepsilon^{\mu\nu}_{\pm}=\varepsilon^{\mu}_{\pm}\varepsilon^{\nu}_{\pm}=\frac{1}{2}(\varepsilon^{\mu\nu}_{11}-\varepsilon^{\mu\nu}_{22}\mp i \varepsilon^{\mu\nu}_{12}\mp i \varepsilon^{\mu\nu}_{21})  
\end{eqnarray}
Now we are going to list some of the identities which are going to be useful in the collinear limit only. Let's specify two momenta $p$ and $q$, which also have similar celestial sphere representations. The corresponding polarization can also be written in the same way as above. The momentum conservation equation $k=p+q$ can be written in the celestial variable as well.
\begin{eqnarray}
    \omega_p(\frac{1+x_1^2}{2},x_1^a,\frac{1-x_1^2}{2})+ \omega_q(\frac{1+x_2^2}{2},x_2^a,\frac{1-x_2^2}{2})=\omega_k ( \frac{1+x_3^2}{2},x_3^a,\frac{1-x_3^2}{2})+ \mathcal{O}(\epsilon)\nonumber\\
    \end{eqnarray}
    here we have introduced $\epsilon= 2 p.q$. In the leading order in the collinear limit, we can justify the above approximation.  Here we will just write the result (see \cite{Jiang:2021csc} for more details)
    \begin{eqnarray}
    \omega_k=\omega_p+\omega_q, \quad \omega_k x_3^a= \omega_p x_1^a+\omega_q x_2^a +\mathcal{O}(\epsilon)\nonumber\\
    x_{13}^a=\frac{\omega_q}{\omega_p+\omega_q} x_{12}^a+\mathcal{O}(\epsilon), \quad  x_{32}^a=\frac{\omega_p}{\omega_p+\omega_q} x_{12}^a+\mathcal{O}(\epsilon), \nonumber\\
    \quad p.\varepsilon_c= \omega_p \hat{p}. \varepsilon_c(k)=\frac{\omega_p \omega_q}{\omega_p+\omega_q} x_{12}^c+\mathcal{O}(\epsilon)
\end{eqnarray}
The contraction of polarization indices can be worked out as well. These are
\begin{eqnarray}
\varepsilon_a(p).q \varepsilon_b(q).\varepsilon_c=-\omega_q x_{12}^a \delta^{bc}, \quad \varepsilon_a(x_i).\varepsilon_b(x_j)=\delta^{ab}, \varepsilon_b(q).p \varepsilon_a(p).\varepsilon_c=\omega_p x_{12}^b \delta^{ac} 
\end{eqnarray}
\section{Graviton OPE from chiral world-sheet}
\label{w symmetry}
In this section, we will find the celestial OPE of two gravitons at the celestial sphere with generic momenta (not necessarily soft). The OPE at the celestial sphere can be obtained as the collinear limit of scattering amplitudes. We will take the two vertex operators in $(-1,-1)$ and $(0,0)$ pictures with momenta $p$ and $q$ respectively which are not necessarily soft. The OPE of these two vertex operators with generic momenta is not very simple to write down. But in the collinear limit ($p||q$), the OPE can be written down \cite{Jiang:2021csc} up to $\mathcal{O}(p.q)$. Once again we will use the HSZ gauge.
\begin{eqnarray}
\int d^2 z_s  V_G^{(0,0)}(z_s^+,z_s^-)V_G^{(-1,-1)}(z_i^+,z_i^-)&&\sim -\frac{1}{4}\int d^2 z_s \mathrm{exp}(\frac{1}{\beta}\frac{p.q\,\, z_{si+}}{z_{si-}})\frac{1}{\beta z_{si-}^2}e^{i (p+q).X}\varepsilon_a^{\mu}\varepsilon_b^{\nu}\nonumber\\
&&(q^{\mu }\varepsilon_c^{\nu}-p^{\nu}\varepsilon_c^{\mu}+\eta^{\mu\nu} p.\varepsilon_c)\varepsilon_{\tilde{a}}^{\tilde{\mu}}\varepsilon_{\tilde{b}}^{\tilde{\nu}}
(q^{\tilde{\mu} }\varepsilon_{\tilde{c}}^{\tilde{\nu}}-p^{\tilde{\nu}}\varepsilon_{\tilde{c}}^{\tilde{\mu}}+\eta^{\tilde{\mu}\tilde{\nu}} p.\varepsilon_{\tilde{c}}) \varepsilon_c.\psi \varepsilon_{\tilde{c}}.\tilde{\psi}e^{-\phi}e^{- \tilde{\phi}}\nonumber\\
\end{eqnarray}
The polarization of gravitons are  $\varepsilon^{\mu\tilde{\mu}}_{a\tilde{a}}(p)=\varepsilon_a^{\mu}(p) \varepsilon_{\tilde{a}}^{\tilde{\mu}}(p)$ and  $\varepsilon^{\mu\tilde{\mu}}_{b\tilde{b}}(q)=\varepsilon_b^{\mu}(q) \varepsilon_{\tilde{b}}^{\tilde{\mu}}(q)$.  After the OPE, we have a resulting graviton with $\varepsilon^{\mu\tilde{\mu}}_{c\tilde{c}}(p+q)=\varepsilon_c^{\mu}(p+q) \varepsilon_{\tilde{c}}^{\tilde{\mu}}(p+q)$ polarization and momenta $p+q$. \\

The world sheet integral in the HSZ gauge can be done as before. The  $z_s^+$ integral produce $\delta(\frac{p.q}{\beta z_{si-}})$. Then  the $z_s^-$ integral will yield $\frac{1}{p.q}$ 
\begin{eqnarray}
\label{w ope}
\int d^2 z_s  V_G^{(0,0)}(z_s^+,z_s^-)V_G^{(-1,-1)}(z_i^+,z_i^-)\sim -\frac{1}{2 p.q}e^{i (p+q).X}\varepsilon_a^{\mu}\varepsilon_b^{\nu}(q^{\mu }\varepsilon_c^{\nu}-p^{\nu}\varepsilon_c^{\mu}+\eta^{\mu\nu} p.\varepsilon_c)\nonumber\\
\varepsilon_{\tilde{a}}^{\tilde{\mu}}\varepsilon_{\tilde{b}}^{\tilde{\nu}}
(q^{\tilde{\mu} }\varepsilon_{\tilde{c}}^{\tilde{\nu}}-p^{\tilde{\nu}}\varepsilon_{\tilde{c}}^{\tilde{\mu}}+\eta^{\tilde{\mu}\tilde{\nu}} p.\varepsilon_{\tilde{c}}) \varepsilon_c.\psi \varepsilon_{\tilde{c}}.\tilde{\psi}e^{-\phi}e^{- \tilde{\phi}}\nonumber\\
\end{eqnarray}
After doing the contraction with polarization we have
\begin{eqnarray}
 =-\frac{1}{2 p.q}e^{i(p+q).X}e^{-\phi}e^{-\tilde{\phi}}\varepsilon_c.\psi \varepsilon_{\tilde{c}}.\tilde{\psi}&&\Big[ \varepsilon_a(p).q \varepsilon_b(q).\varepsilon_c-\varepsilon_b(q).p \varepsilon_a(p).\varepsilon_c+\varepsilon_a.\varepsilon_b p.\varepsilon_c\Big]\nonumber\\
 &&\Big[ \varepsilon_{\tilde{a}}(p).q \varepsilon_{\tilde{b}}(q).\varepsilon_{\tilde{c}}-\varepsilon_{\tilde{b}}(q).p \varepsilon_{\tilde{a}}(p).\varepsilon_{\tilde{c}}+\varepsilon_{\tilde{a}}.\varepsilon_{\tilde{b}}p.\varepsilon_{\tilde{c}}\Big]\nonumber\\
\end{eqnarray}
Next, we need to do the Mellin transformation with respect to $\omega_p$ and $\omega_q$. 
\begin{eqnarray}
    =\int d\omega_p d \omega_q \omega_p^{\Delta_1-1}\omega_q^{\Delta_2-1}\frac{-1}{2 \omega_p \omega_q \hat{p}.\hat{q}}\Bigg(&&\Big[ \omega_q \varepsilon_a(p).\hat{q} \varepsilon_b(q).\varepsilon_c-\omega_p \varepsilon_b(q).\hat{p} \varepsilon_a(p).\varepsilon_c+\omega_p\varepsilon_a.\varepsilon_b \hat{p}.\varepsilon_c\Big]\nonumber\\
    &&\Big[ \omega_q\varepsilon_{\tilde{a}}(p).\hat{q} \varepsilon_{\tilde{b}}(q).\varepsilon_{\tilde{c}}-\omega_p\varepsilon_{\tilde{b}}(q).\hat{p} \varepsilon_{\tilde{a}}(p).\varepsilon_{\tilde{c}}+\omega_p\varepsilon_{\tilde{a}}.\varepsilon_{\tilde{b}}\hat{p}.\varepsilon_{\tilde{c}}\Big] V_{G,p+q}^{(-1,-1)}(z_i^+,z_i^-)\Bigg)\nonumber\\
\end{eqnarray}
We will use the following formula to do the Mellin transformation 
\begin{eqnarray}
    \int_{0}^{\infty} d\omega_1 \omega_1^{\Delta_1-1}\int_{0}^{\infty} d\omega_2\omega_2^{\Delta_2-1} \omega_1^{\alpha}\omega_2^{\beta} (\omega_1+\omega_2)^{\gamma}F(\omega_1+\omega_2)=B(\Delta_1+\alpha,\Delta_2+\beta) \int_{0}^{\infty} \omega^{\Delta_p-1}F(\omega)\nonumber\\
\end{eqnarray}
here $\Delta_p=\Delta_1+\Delta_2+\alpha+\beta+\gamma$.
Hence the terms can be simplified \footnote{It is very crucial to note that we are working in the strict collinear limit. This is the reason these contractions can be simplified.} following the identities in appendix \ref{identities}.
\begin{eqnarray}
\label{w ope final}
  && =\int d\omega_{1} d \omega_2  \omega_{1}^{\Delta_{1}-1}\omega_{2}^{\Delta_{2}-1}\frac{1}{ \omega_p \omega_q x_{12}^2}\Big[ -\omega_q x_{12}^a \delta^{bc}- \omega_p x_{12}^b \delta^{ac}+\frac{\omega_p \omega_q}{\omega_p+\omega_q}x_{12}^c\delta^{ab}\Big]\nonumber\\
  &&\Big[ -\omega_q x_{12}^{\tilde{a}} \delta^{\tilde{b}\tilde{c}}- \omega_p x_{12}^{\tilde{b}} \delta^{\tilde{a}\tilde{c}}+\frac{\omega_p \omega_q}{\omega_p+\omega_q}x_{12}^{\tilde{c}}\delta^{\tilde{a}\tilde{b}}\Big] V_{G,p+q}^{(-1,-1)}(z_i^+,z_i^-)\nonumber\\
 &&  =\frac{1}{x_{12}^2}\Bigg[B(\Delta_1-1,\Delta_2+1) \delta^{bc}\delta^{\tilde{b}\tilde{c}}x_{12}^a x_{12}^{\tilde{a}}+B(\Delta_1,\Delta_2) \delta^{ac}\delta^{\tilde{b}\tilde{c}}x_{12}^b x_{12}^{\tilde{a}}
   +B(\Delta_1,\Delta_2) \delta^{bc}\delta^{\tilde{a}\tilde{c}}x_{12}^a x_{12}^{\tilde{b}}\nonumber\\
   &&+B(\Delta_1+1,\Delta_2-1) \delta^{ac}\delta^{\tilde{a}\tilde{c}}x_{12}^b x_{12}^{\tilde{b}}+B(\Delta_1+1,\Delta_2+1)x_{12}^c x_{12}^{\tilde{c}} \delta^{ab}\delta^{\tilde{a}\tilde{b}}-B(\Delta_1,\Delta_2+1)x_{12}^c x_{12}^{\tilde{a}} \delta^{ab}\delta^{\tilde{b}\tilde{c}}\nonumber\\
   &&-B(\Delta_1+1,\Delta_2)x_{12}^c x_{12}^{\tilde{b}}\delta^{ab}\delta^{\tilde{a}\tilde{c}}-B(\Delta_1,\Delta_2+1)x_{12}^a x_{12}^{\tilde{c}} \delta^{bc}\delta^{\tilde{a}\tilde{b}}-B(\Delta_1+1,\Delta_2)x_{12}^b x_{12}^{\tilde{c}}\delta^{ac}\delta^{\tilde{a}\tilde{b}}\Bigg]\nonumber\\
 &&  \int d \omega \omega^{\Delta_1+\Delta_2-1} V_{G}^{(-1,-1)}(z_i^+,z_i^-)\nonumber\\
 &&=\mathrm{OPE \,\,\, coefficients} \,\,\, \times V_{G,\Delta_1+\Delta_2}
\end{eqnarray}

We can write the above OPE in any dimension. Here we will write this in the $4d$ kinematics (see appendix \ref{identities}).  \\

 The graviton operator in helicity basis is written as
 \begin{equation}
     O_{\Delta,\pm 2}(\mathrm{z},\bar{\mathrm{z}})=\frac{1}{2}(O_{\Delta_1,11}(x)-O_{\Delta_1,22}(x)\mp i O_{\Delta_1,12}(x)\mp i O_{\Delta_1,21}(x))
 \end{equation}
  After doing the OPE of the corresponding operator and writing it on the helicity basis we have the graviton OPE as
\begin{eqnarray}
\label{w ope final 4d}
    \mathcal{O}_{\Delta_1,+2}(\mathrm{z}_1,\bar{\mathrm{z}}_1) \mathcal{O}_{\Delta_2,+2}(\mathrm{z}_2,\bar{\mathrm{z}}_2)&\sim& - \frac{\bar{\mathrm{z}}_{12}}{\mathrm{z}_{12}}B(\Delta_1-1,\Delta_2-1) \mathcal{O}_{\Delta_1+\Delta_2,+2}+...\nonumber\\
     \mathcal{O}_{\Delta_1,+2}(\mathrm{z}_1,\bar{\mathrm{z}}_1) \mathcal{O}_{\Delta_2,-2}(\mathrm{z}_2,\bar{\mathrm{z}}_2)&\sim& - \frac{\bar{\mathrm{z}}_{12}}{\mathrm{z}_{12}}B(\Delta_1-1,\Delta_2+3) \mathcal{O}_{\Delta_1+\Delta_2,-2}
     -\frac{\mathrm{z}_{12}}{\bar{\mathrm{z}}_{12}}B(\Delta_1+3,\Delta_2-1) \mathcal{O}_{\Delta_1+\Delta_2,+2}\nonumber\\
      \mathcal{O}_{\Delta_1,-2}(\mathrm{z}_1,\bar{\mathrm{z}}_1) \mathcal{O}_{\Delta_2,-2}(\mathrm{z}_2,\bar{\mathrm{z}}_2)&\sim& - \frac{\mathrm{z}_{12}}{\bar{\mathrm{z}}_{12}}B(\Delta_1-1,\Delta_2-1) \mathcal{O}_{\Delta_1+\Delta_2,-2}+...\nonumber\\
\end{eqnarray}
This matches with the results in the literature \cite{Guevara:2019ypd,Guevara:2021abz,Pate:2019lpp,Strominger:2021lvk,Jiang:2021csc}. This concludes our discussion of graviton OPE. The OPE coefficients $B(\Delta_1-1,\Delta_2-1)$ has poles whenever $\Delta_1,\Delta_2\in (1,0,-1,-2\cdots)$. One can define the conformal soft theorem and their OPE. The OPE of conformal operators generates the $w_{1+\infty}$ algebra \cite{Guevara:2021abz,Strominger:2021lvk,Jiang:2021csc}.

\bibliographystyle{JHEP}
\bibliography{ref}
\end{document}